\documentclass[10pt,twoside]{article}
\usepackage{amssymb}
\usepackage{amsmath}
\usepackage{epsfig}
\usepackage{color}
\font\tenbf=cmbx9

\font\twelbf=cmbx12
\font\tenrm=cmr9

\font\tenit=cmti9

\newcommand{\lle}{\mbox{$\langle$}}
\newcommand{\rle}{\mbox{$\rangle$}}

\newcommand{\bfvarepsilon}{\mbox{\boldmath$\varepsilon$}}

\newcommand{\bfsigma}{\mbox{\boldmath$\sigma$}}

\newcommand{\bfsi}{\mbox{\boldmath$\sigma$}}
\newcommand{\bfep}{\mbox{\boldmath$\varepsilon$}}

\newcommand{\bfn}{\mbox{\boldmath$\bf n$}}

\newcommand{\bft}{\mbox{\boldmath$\bf t$}}

\newcommand{\bfu}{\mbox{\boldmath$\bf u$}}

\newcommand{\bfx}{\mbox{\boldmath$\bf x$}}
\newcommand{\bfy}{\mbox{\boldmath$\bf y$}}

\newcommand{\bfB}{\mbox{\boldmath$\bf B$}}

\newcommand{\bfD}{\mbox{\boldmath$\bf D$}}

\newcommand{\bfI}{\mbox{\boldmath$\bf I$}}

\newcommand{\bfL}{\mbox{\boldmath$\bf L$}}
\newcommand{\bfN}{\mbox{\boldmath$\bf N$}}

\newcommand{\bfQ}{\mbox{\boldmath$\bf Q$}}

\newcommand{\bfY}{\mbox{\boldmath$\bf Y$}}
\newcommand{\bfZ}{\mbox{\boldmath$\bf Z$}}

\newcommand{\bfR}{\mbox{\boldmath$\bf R$}}

\newcommand{\bfM}{\mbox{\boldmath$\bf M$}}
\newcommand{\bfT}{\mbox{\boldmath$\bf T$}}
\newcommand{\bfS}{\mbox{\boldmath$\bf S$}}

\newcommand{\bfdel}{\mbox{\boldmath$\delta$}}

\newcommand{\bfbe}{\mbox{\boldmath$\beta$}}

\newcommand{\bfGa}{\mbox{\boldmath$\Gamma$}}

\newcommand{\bfeta}{\mbox{\boldmath$\eta$}}

\newcommand{\BB}{\begin{equation}}
\newcommand{\EE}{\end{equation}}

\newcommand{\BBEQ}{\begin{eqnarray}}
\newcommand{\EEEQ}{\end{eqnarray}}
\begin{document}

\vspace{18pt}
 \centerline{\twelbf
Modeling of random bimodal structures of }


\vspace{2pt}

 \centerline{\twelbf composites (application to solid propellants):}
 \vspace{2pt}

 \centerline{\twelbf II. Estimation of effective elastic moduli}

\vspace{6pt}

\vspace{10pt}  \centerline{\bf V.A. Buryachenko$^{1,2)}$\footnote{\noindent \tenrm
$^{1)}$Current address: Micromechanics \& Composites LLC, 2520 Hingham Lane, Dayton, OH 45459, USA\\
$^{2)}$IllinoisRocstar LLC, 60 Hazelwood Drive, Champaign, IL 61820, USA}}

\vspace{2pt}


\vspace{4pt}


\noindent {\baselineskip=9pt
{\bf Abstract:}
We consider a linearly elastic composite medium, which consists of a homogeneous matrix
containing a statistically homogeneous set of multimodal spherical inclusions modeling
 the morphology of heterogeneous solid propellants (HSP).
Estimates of effective elastic moduli are performed using the multiparticle
effective field method (MEFM) directly taking into account the interaction of
different inclusions. Because of this, the effective elastic moduli of the HSP
evaluated by the MEFM are sensitive to both the relative size of the inclusions
(i.e., their multimodal nature) and the radial distribution functions (RDFs) estimated from experimental
data, as well as from the ensembles generated by the method proposed. Moreover,
the detected increased stress concentrator factors at the larger particles in
comparison with smaller particles in bimodal structures is critical
for any nonlinear localized
phenomena for HSPs such as onset of yielding, failure initiation, damage accumulation, ignition, and detonation.}
\vspace{-3pt}

\noindent
{\bf Keywords:}  Microstructures,  random structures, inhomogeneous material,
 effective elastic moduli

\smallskip

\medskip
\noindent{\bf 1. Introduction}
\medskip

Heterogeneous solid propellants (HSPs) commonly used in aerospace propulsion are likely to play an important role for as long as rockets are built.
The typical HSPs consist
 of multimodal spherical particles of solid oxidizer (ammonium perchlorate, AP) of order 1-100 $\mu$m diam embedded in
a rubbery fuel binder such as hydroxi--terminated--polybutadiene (HTPB) or
polybutadiene acrylonitrile (PBAN) (see, e.g. Sutton and Biblarz, 2003).
 Designers of rockets are concerned with a
number of propellant--related issues, including the burning rate, the thermal and
mechanical properties of the propellants, and, for metallized propellants, the
behavior of the metal particles at the surface, including agglomeration
(see Massa, Jackson and Short, 2003; Matou\v{s} and Geubelle, 2006; Matou\v{s} {\color{black}{\it et al.}}, 2007;
Maggi {\it et al.}, 2008;  {\color{black}Xu, Aravas and Sofronis}, 2008).

Both numerical and experimental investigations of random structures are intended for prediction of overall mechanical and physical properties of these media. Suspensions probably is one of the most extensively studied
area of highly packed particulate microheterogeneous media. Numerous authors minimize the non-hydrodynamic
factors considering the low shear rate Newtonian limit that is most interesting from the point of view of analogy with composite materials. In simplest cases, one assumes that at the plotting
of the relative effective viscosity $\eta^*/ \eta^{(0)}$
of a suspension versus the adjustable volume fraction
$c/c_m$, viscosity versus concentration plots collapse onto one curve (see for
references Stickel and Powell, 2005; Arefinia and Shojaei, 2006): $\eta^*/ \eta^{(0)}=[1-f(c^{(1)},\ldots,c^{(n)})]^{-Ac_m}$. He and Ekere (2001) (see also Chong, Christiansen and Baer,  1971; Shapiro and Probstein, 1992; Patlazhan, 1993; D'Haene and Mewis, 1994) proposed semi-analytical viscosity model of {\color{black}noncolloidal} bimodal suspension where $c_m$ depends upon both the size ratio $\lambda$ of large and small particles and the volume fraction $\xi$ of large particles.
In a limiting case $\lambda\to \infty$, the mentioned models are reduced to
 a fruitful alternative by Farris (1968) (see also Probstein, Sengun and Tseng, 1994; Greenwood, Luckham and Gregory,  1998; Zaman and Moudgil, 1998)
 based on purely geometric arguments.
 Assuming that the suspension of the fine particles can be treated as the suspending
media for the large particles, Farris (1968) simulated the full bimodal mixture of particles by mapping it onto an effective one-modal system which can be simulated more easily. In this model, the fine
and coarse fractions are assumed to behave independently of each other and
 the viscosity of the mixture of smaller fractions is the medium
 viscosity for the next largest fraction: $\eta^*/\eta^{(0)}=f(c^{(1)})\cdot f(c^{(2)})\cdot\ldots \cdot f(c^{(n)})$
 ($a^{(1)}\ll a^{(2)} \ll \ldots \ll a^{(n)}$).
 A similar approach (under the name a two-step homogenization technique) is well known (even if the name Farris (1968) is not mentioned) in the engineering material science of both the composites with the large size ratio $\lambda$ of fillers and agglomerated composite materials (for references see, e.g., Buryachenko, 2007a; Enikolopyan, {\it et al.,} 1990; Clements and  Mas, 2004).

Composites with a wide range of bimodal inclusion concentrations
are less well understood and call for further investigation.
So, a standard so-called multi-particle unit cell method of computational micromechanics
is based on periodization
of generated random media (see for references Gusev, Lusti and Hine, 2002; Lusti and Gusev, 2004; B\"ohm, Han and Eckschlager, 2006; Buryachenko, 2007a) reducing to the modeling of periodic boundary
conditions at the meso cell containing just a few tens of inclusions that eliminates {\color{black}both} the boundary and size effects in a windowing approach (see, e.g., Liu {\it et al.}, 2009).
The generation process is based upon Monte Carlo simulation method wherein the particles are generated from a certain aggregate size distribution and then randomly placed into the meso cell in such a way that there is no intersection between the particles. Then, the generated mesostructure models are to be meshed
for computational testing of the composite with the finite element method (FEM) in order to estimate of the macroscopic response [the use of the boundary element method (Chen and Papathanasiou, 2004) and multipole method (Hatch and Davis, 2006; Kushch, Sevostianov and Mishnaevsky, 2008) {\color{black}is also known}]. In so doing, the pack generations are based on either the hard core model simulation (Segurado and Llorca, 2002; Wissler {\it et al.}, 2003;
H\"afner, {\it et al.}, 2006; Wriggers and Moftah, 2006; Kushch, Sevostianov, and Mishnaevsky, 2008; Tawerghi and Yi, 2009) or the collective rearrangement model (CRM) destined for creation of the
close packing (Stroeven, Askes, Sluys, 2004; Annapragada, Sun and Garimella, 2007; Matou\v{s}, {\it et al.}, 2007). Disadvantages of these random generation methods for composites with a wide range of particle concentration were considered by Buryachenko {\it et al.} (2003). The advances in imaging techniques [e.g., computed tomography (CT), scanning electron microscopy (SEM), and magnetic resonance imaging (MRI)] require powerful computational methods for numerical analysis of CMs.
To simplify the technology of finite element mesh generation for particle reinforced
material (e.g., HSP), enrichment techniques is used to account for the material interfaces in the framework of extended FEM (XFEM, see for references Fries and Belytschko, 2010;
Hiriyur, Waisman, and Deodatis, 2011), first introduced by Belytschko and Black (1999) as a solution to the remeshing issue for crack propagation.  The geometry of material distribution is described by level set function, which allows one to model the internal boundaries of the microstructure without the adaptation of the mesh (Du, Ying and Jiang, 2010; Legrain {\it et al.}, 2011). Some alternative approaches  based on a real structure can be done mainly in two different ways. {\color{black}The first one can } directly import the experimentally obtained real structure into FE software (see for references Ghosh, 2011; Legrain {\it et al.}, 2011).
The second approach based on the concept of Statistically Equivalent Periodic Unit Cell (SEPUC) generates the model structures
that have similar statistical functions as the experimentally obtained ones with a subsequent substitution of found descriptor into analytical micromechanics models (see, e.g.,
Povirk, 1995; Matou\v{s}, Lep\v{s}, Zeman and \v{S}ejnoha, 2000; Zeman and \v{S}ejnoha,  2001;
Borb\'ely,  Kenesei and Biermann, 2006; Lee, Gillman and Matou\v{s}, 2011).

The modeling of composites with bimodal spherical inclusions by the methods of analytical micromechanics
(see for references Nemat-Nasser and Hori, 1993; Torquato, 2002; Milton, 2002;
Buryachenko, 2007a) has been developed less
and {\color{black}necessitates} some additional consideration. It is explained by a fact that the most popular methods of
analytical micromechanics are essentially one--particle ones.
First of all, this will be true for the variants of the effective medium method by Kr\"oner (1958) and by Hill (1965) and the mean field method (Mori and Tanaka, 1973; Benveniste, 1987). Recently a new method has become known, namely the
multiparticle effective field method (MEFM) was put forward and developed by one of the authors (see for references Buryachenko, 2007a). The MEFM is based on the theory of functions of random variables and {\color{black}the Green}
functions. MEFM does not make use of a number
of hypotheses which form the basis of the traditional one-particle methods.
 It is obvious that the classical methods mentioned above do not allow for direct binary interaction of inclusions, and, because of this, these methods are invariant to the size distribution of inclusions. Contrastingly, the MEFM has qualitative benefits following immediately from the consideration
of multiparticle interactions and manifesting by a sensitivity of the MEFM's estimations to the size distribution of heterogeneities.

So called discrete structural models alternatively taking binary interactions of inclusions into account are worthy of notice (see, e.g.,
Zgaevskii, 1977; Garishin, 1997; Garishin and Moshev, 2002; Yanovsky and  Zgaevskii, 2004).
These methods were initiated by Chen and Acrivos (1978) who showed that the elastic energy stored in the matrix layer between the spheres is many times greater than that in the rest of the surrounding matrix. This matrix layer can be considered as the elastic
rod or spring transferring the interaction force between the spheres
that makes it possible to use the principle of physical discretization developed by Absi and Prager (1975).
A composite material {\color{black} is} modeled
as an elastic network which seems to be much simpler for mathematical treatment than the original continuum mechanics representation.
The fundamental limitations of the method are a simple composite structure and the growing of an approximate error of a spring model with both the growing of inter-particle distances and decreasing of elastic mismatch of constitutives.

The paper is organized as follow.
In section 2 we present the basic field equations of linear elasticity, notations, and statistical description of the composite microstructure.
In Section 3 we recall the basic concepts defining some classical method of micromechanics and present explicit formula for both effective elastic moduli and stress concentrator factor for composites with bimodal distribution of spherical particles {\color{black} and the radial distribution functions (RDFs) evaluated in an accompanying paper by Buryachenko,  Jackson  and Amadio (2012), henceforth referred to as (I).}
In Section 4 it is detected that the effective elastic moduli of the HSP evaluated by the MEFM are very sensitive (that is confirmed by comparison with available experimental data) to both the relative size of inclusions (i.e. their multimodal nature) and the RDFs. Moreover,
 one detects increased stress concentrator factors at the big particles in comparison with one for the small particles in bimodal structures that is critical for any nonlinear localized
phenomena such as onset of yielding, failure initiation, damage accumulation, and others. A wide class of prospective problems for the modeling of HSPs is considered in Section 5.

\bigskip
\noindent
{\bf 2. Preliminaries}

\medskip
\setcounter{equation}{0}
Let a linear elastic infinite body $R^d$
{contains} an open bounded domain $w\subset R^d$ (window of observation)
with a boundary $\Gamma$ and with an indicator
function $W$ and space dimensionality $d$
($d=2$ and $d=3$ for 2-$D$ and 3-$D$ problems, respectively).
The domain $w$ contains a homogeneous matrix $v^{(0)}$ and
a random finite set $X=(v_i)$ $(i=1,\ldots, N(w))$ of inclusions
$v_i$ with centers $\bfx_i$ and with characteristic
functions $V_i(\bfy)$ equals one {\color{black}if} $\bfy\in v_i$ and zero otherwise
and bounded by the spherical surfaces
$\Gamma_i^{(1)}=\{\bfy:\ |\bfy-\bfx_i|=a^{(1)}\}$ and $\Gamma_i^{(2)}=\{\bfy:\ |\bfy-\bfx_i|=a^{(2)}\}$ of two radii
$a^{(1)}$ and $a^{(2)}$, respectively; $v^{(k)}=\cup v_i^{(k)}$ $(i=1,2,\ldots, \ k=1,2)$.

{\color{black} The local strain tensor
  $\bfvarepsilon$  is related to the displacements
$\bf u$ via the linearized strain--displacement equation
 \BB\bfvarepsilon ={1\over 2}[\nabla \otimes {\bf u}+(\nabla \otimes
{\bf u})^{\top}].
\EE
            Here $\otimes$ denotes
tensor product,
             and $(.)^{\top}$  denotes matrix transposition. The stress tensor
            $\bfsi $ satisfies the equilibrium equation  (no body forces
acting):
         \BB\nabla \bfsi =0.
\EE
  Stresses and strains are related to each other  via the constitutive equations
\BB
\bfsi ({\bf x})={\bf L(x)}\bfvarepsilon ({\bf x}) \quad
 {\rm or }\quad \bfvarepsilon ({\bf x})={\bf M(x)}\bfsi({\bf x}).
\EE
 ${\bf L(x)}$ and ${\bf M(x) \equiv L(x) }^{-1}$
are the known phase stiffness and  compliance
fourth-order tensors, and the common notations for scalar products
have  been employed: ${\bf L}\bfvarepsilon =L_{ijkl}\varepsilon_{kl}$.
 In particular, for isotropic constituents the local stiffness
tensor $\bfL(\bfx)$ is given in terms of the local bulk modulus $k(\bfx)$ and the local
shear modulus $\mu(\bfx)$:
\BB
\bfL(\bfx)=\Big(dk(\bfx),2\mu(\bfx)\Big)\equiv
dk(\bfx)\bfN_1+2\mu(\bfx)\bfN_2,\quad \bfbe(\bfx)=\beta_0(\bfx)
\bfdel,
\EE
where in component form
\BB
N_{1|ijkl}={1\over d}\delta_{ij}\delta_{kl},\ N_{2|ijkl}={1\over 2}
(\delta_{ik}\delta_{jl}+\delta_{il}\delta_{jk})-{1\over d}\delta_{ij}
\delta_{kl}
\EE
and $\delta_{ij}$ is the Kronecker delta; $i,j,k,l=1,2,\ldots,d;$ $d=2,3$.

 All
tensors ${\bf f}$ $({\bf f=L, M})$ of material properties are decomposed
          as ${\bf f\equiv f}^{c}+{\bf f}_1({\bf x})$.
          The subscript 1 denotes a
jump of the corresponding quantity (e.g. of the material tensor).
          ${\bf f}$ is assumed
to be constant in the matrix $v^{(0)}=w\backslash     v$
       and inside the     inclusions:
${\bf f(x)}={\bf f}^{(0)}$  for ${\bf x}\in v^{(0)}$, and
               ${\bf f(x)}=  {\bf f}^{(0)}+{\bf f}^{(k)}_1$ for ${\bf x}\in v^{(k)}$.
Here and in the following the upper index $^{(k)}$ numbers the
            components and the lower index $i$ numbers the individual
inclusions; $ v\equiv \cup v^{(k)}\equiv\cup v_i,
\ V(\bfx)=\sum V^{(k)}=\sum V_i(\bfx)$, and $V^{(k)}(\bfx)$ is a
characteristic function of $v^{(k)}$, equal to 1 if
$\bfx\in v^{(k)}$ and 0 otherwise, $(k=1,2;
\ i=1,2,\ldots)$.

                We assume that the phases are perfectly bonded, so that
            the displacements and  the traction components of the stresses
            are continuous across the interphase boundaries.
For the mesodomain $\bfx\in
w$ we take either uniform  displacement or traction
boundary conditions
\BBEQ
\bfu(\bfx)&=&\bfep^0\bfx,\\
\bfsi(\bfx)\bfn(\bfx)&=&\bfsi^0\bfn(\bfx)=\bft(\bfx),
\EEEQ
respectively, where ${\bf t(x)}$ is the traction vector at
the external boundary
$\partial w$, ${\bf n}$ is its unit outward normal, and $\bfep^0$ and
$\bfsigma^0$ are
the mesoscopic strain and stress tensors, i.e. a
given constant symmetric tensor.

All the random quantities  under
discussion are  statistically     homogeneous,
and, hence, the ensemble averaging could be replaced by volume
averaging
\BB
\langle (.) \rangle =\overline w^{-1}\int (.)W({\bf x})\,d{\bf x},\quad
 \langle (.) \rangle ^{(k)} =[\overline v^{(k)}]
^{-1}\int (.)V^{(k)}({\bf x})\,d{\bf x}, 
\EE
where the bar appearing  above the region represents its measure, e.g.
$\overline v\equiv {\rm mes}$ $v$. The average over component $v^{(k)}$
agrees with the ensemble average over an individual inclusion $v_i\in v^{(k)}
 \ \ (i=1,2, \ldots ): \ \  \langle (.) \rangle _i= \langle (.) \rangle ^{(k)}$.
The notation  $ \langle (.) \rangle _i({\bf x})$ at $ {\bf x}\in v_i\subset v^{k}$ means
the average  over an  ensemble
realization of surrounding inclusions (but not over the volume $v_i$ of a
particular  inclusion, in contrast to $ \langle (.) \rangle _i$).
For the description of the random statistically homogeneous structure
of a composite material, let us introduce
 a conditional probability density $\varphi (v_j,{\bf x}_j
\vert; v_i,{\bf x}_i)$,
which is a probability density to find the
$j$-th inclusion  with the center ${\bf x}_j$ in the domain
$v_j$ with fixed inclusion  $v_i$ with the centers ${\bf x}_i$, and ${\bf x}_j\neq {\bf x}_i$.
Of course, $\varphi(v_j, {\bf x}_j\vert ;v_i,{\bf x}_i)$
for values of ${\bf x}_j$ lying
inside the ``excluded volumes'' (called also the
correlation hole) $\cup v^0_{ji}$, where $v^0_{ji}\supset v_i$ with characteristic
functions
$V_{0m}$ (since inclusions cannot overlap), and
 $\varphi(v_j, {\bf x}_j\vert ;v_i,{\bf x}_i)\to \varphi(v_j,
{\bf x}_j)$ at $\vert {\bf x}_j-{\bf x}_i\vert\to \infty$
(since no long-range order is assumed);
it is assumed that
$v^0_{ji}\equiv v_j^0$ are the circles with the
radii $2a;\ j,i=1,2\ldots)$.
Only if the pair distribution
  function $g_{km}({\bf x}_j-{\bf x}_i)\equiv
  \varphi(v_j, {\bf x}_j\vert; v_i,{\bf x}_i)/n^{(m)}$  ($v_i\in v^{(k)},\ v_j\in v^{(m)},\ k,m=1,2$)
  depends on $\vert {\bf x}_j-{\bf x}_i\vert$  it is called the
  radial distribution function.
$\varphi (v_i,{\bf x}_i)$ is a
number density $n^{(k)}$ of component $v^{(k)}\ni v_i$  and $c^{(k)}$ is the
concentration, i.e. volume fraction, of the component $v^{(k)}$:
$ c^{(k)}=\langle V^{(k)}\rangle =\overline v_in^{(k)}$ ($k=1,2;
\  i=1,2,\ldots),\quad
c^{(0)}=1-\langle V\rangle,$  $c=\lle V\rle$.  }

\bigskip
\noindent {\bf 3 Effective elastic moduli and stress concentrator factors}
\medskip

We will slightly modify the basic assumptions and the final
formula of the multiparticle effective field method (MEFM) for estimation of effective
elastic moduli of composites with polydisperse particles. This case was not considered by Buryachenko (2007a) where a detailed discussion
and numerous references for this and related methods can be found.

Substituting (3) and  (1) into
the equilibrium equation (2), we obtain a differential
equation with respect to the displacement $\bfu$ which
may be reduced to  a symmetrized integral form for the stresses
\BB
\bfsi ({\bf x})=
\langle \bfsi\rangle
+\int {\bf \Gamma} (\bfx-\bfy)
[\bfeta({\bf y})
-\langle \bfeta\rangle ] 
 d{\bf y},
\EE
where the tensor $\bfeta({\bf y})={\bf M}_1({\bf y)}\bfsi({\bf y})$ is called the strain polarization tensor
and is simply a notational convenience (see
Willis, 1981).
The integral operator kernel
\BB
{\bf \Gamma} (\bfx-\bfy)= -{\bf L}^{(0)}
\left[{{\bf I}\delta ({\bf x-y})+
{\bf U}({\bf x-y}){\bf L}^{(0)}}\right], 
\EE
is the even homogeneous generalized functions of the order
$-d$ defined by the second derivative of the Green tensor ${\bf G}$:
\BBEQ
 \!\!\!\!U_{ijkl}(\bfx)\!\!\!\!&=&\!\!\!\!\big[\nabla_j\nabla_l G_{ik}(\bfx)\big]_{(ij)(kl)}, 
\EEEQ
where the notation indicates symmetrization on $(ij)$
and $(kl)$, ${\bf I}$ is a unit fourth-order  tensor, and
${\bf G}$ is the infinite-homogeneous-body Green's function of the
Navier equation with an elastic modulus tensor ${\bf L}^{(0)}$,
\BB
\nabla \left\lbrace{{\bf L}^{(0)}
\left[{\nabla \otimes {\bf G}(\bfx)+
 (\nabla\otimes{\bf G}(\bfx))^{\top}}\right]/2}\right\rbrace =-\bfdel \delta
 ({\bf x}), 
\EE
vanishing at infinity ($|\bfx|\to\infty$),
$\delta ({\bf x})$ is the Dirac delta function, $\bfdel$ is
the unit second order tensor.
The equation (12) is valid for the domain $\bfx\in w$ containing statistically large number of inclusions except when the distance of $\bfx$ from the boundary $\partial w$ of the body $w$ is of order of a correlation length of
the microstructure (see for details Willis, 1981; Buryachenko, 2007a).

Both the effective compliance ${\bf M}^*$ and the effective moduli $\bfL^*=(\bfM^*)^{-1}$
appearing in the overall constitutive relation
\BB
\langle \bfep \rangle =\bfM^* \langle \bfsi \rangle , \ \ \langle \bfsi \rangle =\bfL^* \langle \bfep \rangle
\EE
can be defined by the general relation
\BB
{\bf M}^*= \langle {\bf MB}^* \rangle ,
\EE
where ${\bf B}^*={\bf B}^*({\bf x})$ is a local stress concentration
tensor obtained under pure mechanical loading
\BB
\bfsigma({\bf x})={\bf B}^*({\bf x})\lle\bfsigma\rle.
\EE

After conditional statistical averaging Eq. {\color{black}(9)}
turns into an infinite system of integral equations.
In order to close and approximately solve this system we now apply the
MEFM hypotheses (see details Byryachenko, 2007a):

{\bf H1)} {\it Each inclusion $v_i$ has a spherical form and
is located
 in the effective field $\overline{\bfsi} _i({\bf y})$ which is homogeneous
over the inclusion $v_i$ and }
\BB
\overline{\bfsi}_i({\bf y})\equiv \overline{\bfsi}_i=\overline{\bfsi}^{(k)} 
\EE
{\it for any $\bfy\in v_i\subset v^{(k)}\ (k=1,2)$.}

Therefore, since ${\bf L(x)}$ is constant inside the spherical inclusion, we have
\BB
\int{\bf \bfGa({\color{black}\bfx-\bfy})}V_i({\bf y}){\bf M}^{(1)}_1({\bf y})
\bfsi({\bf y})d{\bf y}=
\bar v_i{\bf T}_i({\color{black}\bfx-\bfx_i}){\bf M}^{(1)}_1
 \bfsi_i, 
\EE
where ${\bf x}\notin v_i$ and $\bfsi _i\equiv
\lle\bfsi({\bf y})V_i({\bf y})\rle_{(i)}$ is {\color{black}the random stress field averaged} over
the volume of the inclusion $v_i$ (but not over the ensemble).
Hereinafter $({\bf y} \in v_j)$
\BB
{\bf T}_i{\bf (x-x}_i)=(\overline v_i)^{-1}\int
{\bfGa\bf (x-y)}V_i({\bf y})d{\bf y},\quad
{\bf T}_{ij}{\bf (x}_i-{\bf x}_j) =\lle{\bf T}_i{\bf (y-x}_i)\rle_j.
\EE
The tensor $\bfT_{ij}(\bfx_i-\bfx_j)$ has an analytical representation for spherical
inclusions of different size in an isotropic matrix (see for references Buryachenko, 2007a), the case of
ellipsoidal inclusions of different sizes and orientations is analyzed by Franciosi and Lebail (2004).

According to hypothesis ${\bf H1}$ and
to Eshelby's (1961) theorem we get
\BB
\bfsi({\bf x})={\bf B}\overline
{\bfsi}_i ({\bf x}),\quad
\overline v_i {\bf M}_1^{(1)}\bfsi({\bf x})
={\bf R}_i\overline {\bfsi}_i ({\bf x}),\quad
\quad {\bf x}\in v_i,
\EE
where
\BB
{\bf B =(I+Q}(v_i){\bf M}_1^{(1)})^{-1}{\rm \color{black}=const.},
\ \ {\bf R}_i=\overline v_i{\bf M}_1^{(1)}{\bf B} 
\EE
and tensor ${\bf Q}(v_i)$ is associated with the
well--known Eshelby tensor by
\BB
\bfS=
 \bfI-\bfM^{(0)}\bfQ(v_i). 
 \EE

Using hypothesis {\bf H1} and the solution (20) for one inclusion, we
get algebraic solution for two inclusions $v_1$ and $v_2$ subjected to the effective field
$\widehat {\bfsi}_{i,j}
 ({\bf x}_j)$:
\BB
\overline {\bfsi}_i ({\bf x})=\bfR_i^{-1}\sum^2_{j=1}{\bf Z}_{ij}\bfR_j
\widehat {\bfsi}_{i,j}
 ({\bf x}_j),\quad {\bf x}\in v_i\ \ (i=1,2). 
\EE
 Here the matrix ${\bf Z}^{-1}$ has the elements ($i,j=1,2$)
\BB
({\bf Z}^{-1})_{ij}={\bf I}\delta_{ij}-(1-\delta_{ij})
{\bf R}_j{\bf T}_{ij}({\bf x}_i-
{\bf x}_j). 
\EE

For termination of the hierarchy of statistical moment equations in everaged Eq. (9), we will use the
closing effective field hypothesis:

{\bf H2)} {\it Each {\color{black}pair} of the inclusions $v_i$ and $v_j$ is located in an
effective field
 $\widehat {\bfsi}_{i,j} ({\bf x})$ (22) and}
\BB
\lle\widehat {\bfsi}_{i,j} ({\bf x})\rle_k = \lle\overline{\bfsi}
 ({\bf x})\rle_k,\ {\bf x}\in v_k \quad (k=i,j). 
\EE

 After estimating average stresses inside the inclusions ($i=1,2$)
 \BB
 \lle\bfsi\rle_i= (c^{(i)})^{-1}(\bfM_1^{(i)})^{-1}\sum^2_{j=1} n^{(j)}{\bf Y}_{ij}{\bf R}_j\lle\bfsi\rle, 
 \EE
the problem of calculating effective properties of statistically homogeneous and statistically isotropic medium becomes trivial and leads (see for details Buryachenko, 2007a) to:
\BBEQ
{\bf M}^*&=&{\bf M}^{(0)}+\sum^2_{i,j=1} {\bf Y}_{ij}{\bf R}_jn^{(j)},
\EEEQ
where the matrix ${\bf Y}^{-1}$ has the following elements
$({\bf Y}^{-1})_{ij}\ (i,j=1,2)$:
\BBEQ
 \!\!\!\!\!\!\!&\!\!\!\!&\!\!\!\!\!\!\!\!\! ({\bf Y}^{-1})_{ij}=
 \delta_{ij}\left[{\bfI-\bfR_i
 \sum^2_{q=1} {n^{(i)}\over n^{(q)}} \!\int {\bf T}_{iq}({\bf x}_i-{\bf x}_q){\bf Z}^{21}_{qi}
 \varphi(v_q,\bfx_q|;v_i,\bfx_i)d{\bf x}_q}\right]\nonumber \\
 \!\!\!\!\!\!\!\!\!\!\! &\!\!\!\!-\!\!\!\!&{\bf R}_i {n^{(i)}\over n^{(j)}}\int\left[{
 {\bf T}_{ij}({\bf x}_i-{\bf x}_j){\bf Z}^{22}_{ji}
  \varphi(v_q,\bfx_q|;v_i,\bfx_i)-n^{(j)}{\bf T}_i({\bf x}_i-
 {\bf x}_j)}\right]d{\bf x}_j,
\EEEQ
where $\bfZ_{qi}^{kl}=(\bfZ_{qi}^{kl})$ ($q,i,k,l=1,2$) are the submatrixes ($6\times 6$) of the matrix $\bfZ_{qi}$ ($12\times 12$).

It should be mentioned that the quasi crystalline approximation by Lax (1952) is equivalent to the assumption
\BB
\bfZ_{ij}=\delta_{ij}\bfI, 
\EE
which reduces Eq. (27)  (for the aligned homogeneous ellipsoidal particles) to
\BBEQ
\!\!\!\!\!\!\!\!\!\! ({\bf Y}^{-1})_{ij}\!\!\!\!&=&\!\!\!\!
 \delta_{ij}\bfI\nonumber \\
 \!\!\!\!&-&\!\!\!\! {n^{(i)}\over n^{(j)}}{\bf R}_i\!\!\!\int\!\!\left[{
 {\bf T}_{ij}({\bf x}_i-{\bf x}_j)
  \varphi(v_j,\bfx_j|;\!v_i,\bfx_i)\!-\!n^{(j)}{\bf T}_i({\bf x}_i-
 {\bf x}_j)}\right]\!d{\bf x}_j,
\EEEQ

Due to a widely used second-order probability function $S_{ij}(r)$ (rather than only RDF), we reproduce some classic results obtained in the framework of the quasi crystalline approximation by Lax (1952) in the form by Willis (1977) (see also Hashin and Shtrikman, 1963; Willis, 1981; Lee, Gillman and Matou\v{s},  2011)
\BBEQ
 \lle\bfsi\rle_i&=&(\bfM_1^{(i)})^{-1}\sum^2_{j=1} \widehat{\bfY}_{ij}{\bf M}_1^{(j)}\lle\bfsi\rle, \\
 {\bf M}^*&=&{\bf M}^{(0)}+\sum^2_{i,j=1} c^{(i)}\widehat{\bf Y}_{ij}{\bf M}_1^{(j)},
\EEEQ
where
\BBEQ
\!\!\!\! (\widehat{\bf Y}^{-1})_{ij}\!=\!
 \delta_{ij}\bfI
 \!-(\!c^{(i)})^{-1}{\bf M}_1^{(i)}\!\!\!\int\!
 {\bfGa}({\bf x}_i-{\bf x}_j)[S_{ij}(\bfx_i-\bfx_j)-c^{(i)}c^{(j)}]
d{\bf x}_j,
\EEEQ

To make further progress, the hypothesis of ``{\it ellipsoidal symmetry}" for the distribution of inclusions attributed to Willis (1977) (see also
Khoroshun, 1974, 1978;  Ponte Castaneda and Willis, 1995) is widely used:

\noindent {\bf Hypothesis H3, ``ellipsoidal symmetry"}.
{\it The conditional probability density function $\varphi (v_{j},{\bf x}_j \mid ;v_{i},{\bf x}_{i})$ depends on $\bfx_j-\bfx_i$ only through the combination $\rho=|({\bf a}^0_{ij})^{-1} ({\bf x}_{j}-{\bf x}_{i})|$}:
\BB
\varphi (v_{j},{\bf x}_j \mid ;v_{i},{\bf x}_{i})
=h(\rho ),
\EE
{\it where the matrix $({\bf a}^0_{ij})^{-1}$ (which is symmetric in the indexes $i$ and
$j$, ${\bf a}^0_{ij}={\bf a}^0_{ji}$)
defines the ellipsoid excluded volume $v^0_{ij}=\{\bfx:\ |({\bf a}^0_{ij})^{-1}\bfx|^2< 1\}$.}

In such a case, the representations of the tensors $\bfY{ij}$ and $\widehat{\bfY}_{ij}$ are simplified
\BBEQ
(\bfY^{-1})_{ij}&=&\delta_{ij}\bfI-(\bfI-\bfB)\bar v_in^{(i)},\\  
(\widehat{\bfY}^{-1})_{ij}&=&\delta_{ij}\bfI- (\delta_{ij}-c^{(j)})\bfM_1^{(i)}\bfQ(v_i^0), 
\EEEQ
and both pairs of Eqs. (25), (26) and Eqs. (30) (31) are reduced to the same formulae
\BBEQ
\lle\bfsi\rle_i\!\!\!&=&\!\!\!\bfB
(c^{(0)}+\bfB c)^{-1}\lle\bfsi\rle, \\ 
\bfM^*\!\!\!&=&\!\!\!\bfM^{(0)}+c(c^{(0)}+\bfB c)^{-1}\bfM_1^{(1)}\bfB ,\  
\EEEQ
coinciding with the estimations by Mori and Tanaka (1973) method (MTM).

The destination of the hypothesis {\bf H3} is directed towards providing conditions for applying  the hypothesis {\bf H1}
(see for details Buryachenko,  2010b, 2011).

The MEFM has a few advantages with respect to the classical one-particle method: 1) Direct binary interaction of heterogeneities (see Eq. (22)); 2) Explicit dependence of effective moduli on the partial RDF (see Eqs. (25)-(27)); 3) MEFM is not invariant with respect to the size distribution of spherical particles (see Eqs. (25)-(27)). In doing so, the classical methods (such as, e.g., Hashin and Shtrikman, 1963; Hill, 1965; Mori and Tanaka, 1973; and Willis, 1981,
see Eqs. (36) and (37)) used the quasi crystalline approximation by Lax (1952) are invariant with respect to both the RDF and the size distribution of spherical heterogeneities.

\bigskip
\noindent {\bf 4 Numerical results}
\medskip

\noindent{\textit{\textbf{ 4.1 Unimodal distribution}}
\medskip

 The RDF $g(r)$ is well
investigated only for identical spherical ($3D$ and $2D$ cases, see for references, e.g., Buryachenko, 2007a)
inclusions with a radius $a$. Two
alternative analytical RDFs of inclusion will be examined for 3D case
\BBEQ
g^S(r)\!\!\!\!&=&\!\!\!\!H(r-2a)\\ 
g^W(r)\!\!\!\!&=&\!\!\!\!H(r-2a)\left[{1+
\Big({2+c\over 2(1-c^2)}-1\Big) \cos ({\pi r\over a})e^{2(2-r/a)}
}\right], 
\EEEQ
where $H$ denotes the Heaviside step function, $r\equiv \vert
{\bf x}_i-{\bf x}_q\vert $ is the
distance between the nonintersecting fixed inclusions $v_i$ and moving one $v_q$,
 $c$ is the volume fraction of particles of the radius $a$.
The formula (38) describes a well-stirred approximation (differing from the RDF for a Poisson distribution by the availability of
``excluded volume" with the center $\bfx_i$ where $g(|\bfx_i-\bfx_q|)\equiv 0$)
while Eq. (39), attributed to Willis (1978), takes into account a neighboring order in the distribution of the inclusions (see for details I).
We will also consider
the popular Percus-Yervik (1958) approximation $g^{P-Y}(r)$ used for construction of Eq. (39): $g^W(2a)=g^{P-Y}(2a)$.

In order to demonstrate the comparison of the available experimental data
with the predicting capability of the proposed method, we will
consider the estimation of the effective elastic moduli $\bfL^*$ (25).
An experimental data set from
the work of Smith (1976) used for validation were obtained for composite materials
consisting of identical glass beads embedded in an epoxy matrix.
Young modulus and Poisson ratio of the particles and the matrix $E^{(1)}=76.0$ GPa, $\nu^{(1)}=0.23$
and $E^{(0)}=3.01$ GPa, $\nu^{(0)}=0.394$, respectively.
As can be seen from
Fig. 1 the use of the approach (37) based on the
   quasi-crystalline approximation (27)
[also called Mori-Tanaka method (MTM, see Mori and Tanaka, 1973; and Benveniste, 1983), curve 4] leads to an underestimate
of the normalized effective Young modulus $E^*/E^{(0)}$ by 1.5 times for c = 0.5 compared with the
experimental data. Much better approximations of
$E^*/E^{(0)}$ are given by the MEFM (23) and (25) with the RDF obtained by the CSM, SCRM, and Willis approximation (39) which are
distinguished one from another by less than 0.5\%. In so doing, the use of the
well-stirred RDF (38) instead of (39) leads to reduction of predicted values $E^*/E^{(0)}$ on 10\%.
Therefore, in the considered case of the elastic mismatch of constituents,
the effective elastic moduli estimated by the MEFM (23), (25) are little
sensitive with respect the RDF taking into account a neighboring order in the
particle distribution. In doing so, a reduction of the proposed approach to the
classical one (27) (which is invariant with respect to any RDFs) leads to
significant underestimation of $E^*/E^{(0)}$.

Let us now demonstrate an application of the theoretical results by considering
an isotropic composite made of an incompressible isotropic matrix, filled with
rigid identical spherical inclusions. This example was chosen deliberately because
it provides the maximum difference between predictions
of
relative effective shear modulus
$\mu^*/\mu^{(0)}$, as estimated by the various methods.
It should be mentioned that in the case being considered, $\mu^*/\mu^{(0)}$ is equivalent to relative change of the Newtonian
viscosity $\eta^*/\eta^{(0)}$ of a suspension of identical rigid spheres, see experimental data by Kreger (1972). In Fig. 2 the
micromechanical
model MEFM (23) and (25) is analyzed for the
effect of choosing different RDFs (see for details I).
As {\color{black}can be} seen, the effective shear moduli can differ by a factor of two or more
depending on the chosen RDF.
It is interesting that the first peak in Willis (39) approximation is lower than the first peak in both the {\color{black} collective rearrangement model (CRM) and shaking collective rearrangement model} (SCRM} simulation, but the RDF (39) decreases slower, and, because of this, the effective shear moduli $\mu^*$ for the RDF (39) {\color{black}larger} than $\mu^*$ corresponded to the CRM and SCRM. In contrast
 to the Willis approximation (39),
PY approximation swiftly decreasing after the first peak leads to the values of $\mu^*$ which are below corresponding $\mu^*$ estimated for both the CRM and SCRM.
Just for estimation of the approximation quality (17I),
the curve 2 in Fig. 2 is evaluated for the RDF predicted from a single SCRM's 

\vspace{-13mm}
\noindent \parbox{7.8cm}{
\hspace{-25mm} \psfig{figure=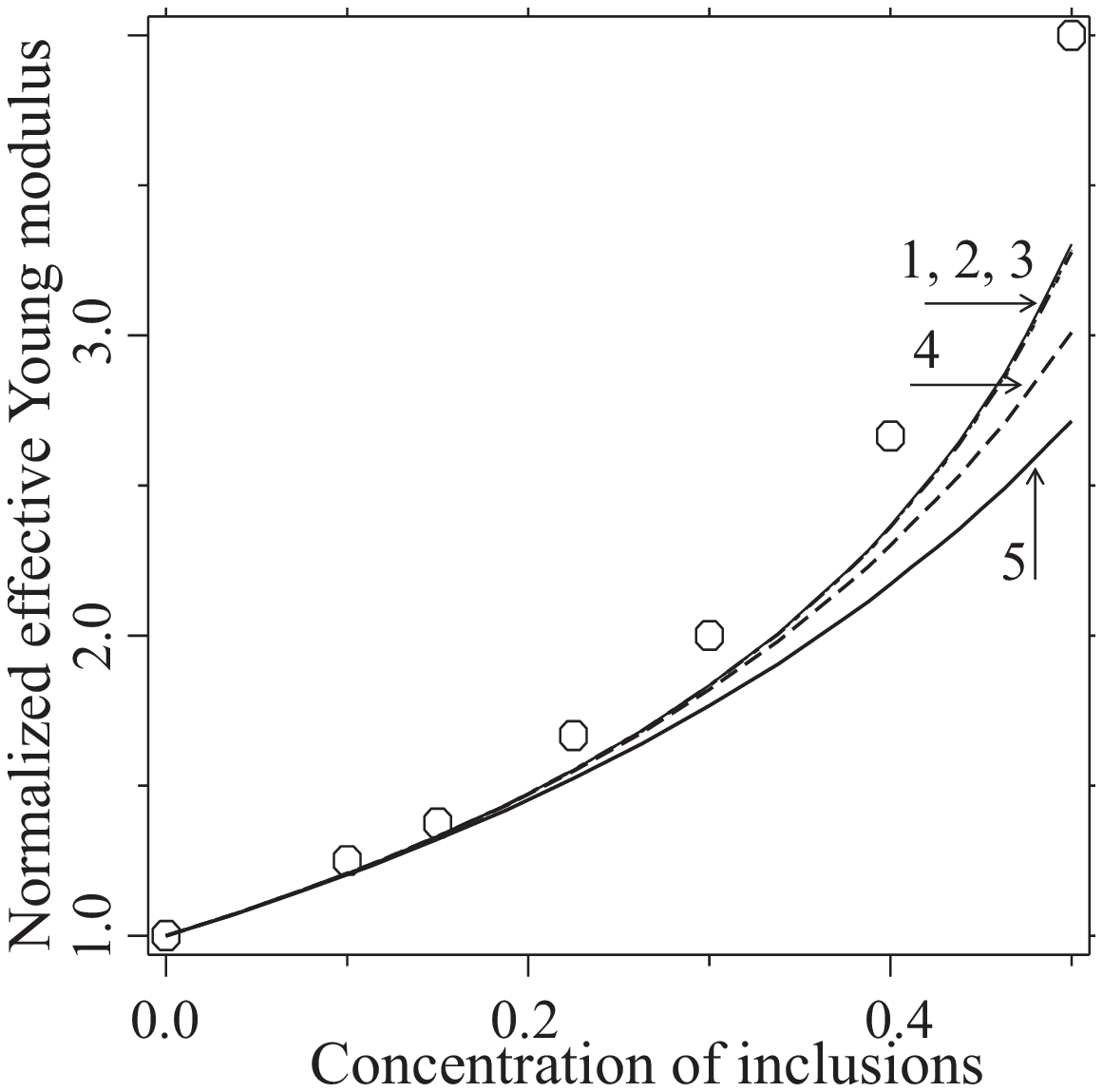, width=12.cm}\\
\vspace{-87mm} \tenrm \baselineskip=8pt

{\rm Fig. 1}:
$E^*(c)/E^{(0)}$ vs $c$ estimated by the MEFM with RDF simulated by CRM (1), SCRM (2), Willis approximation (3), well-stirred approximation (4), and by the MTM (5); $\circ$ - experimental data by Smith [105]}
 \ \ \noindent \parbox{7.8cm}{
\vspace{-8mm} \hspace{-20mm}
\psfig{figure=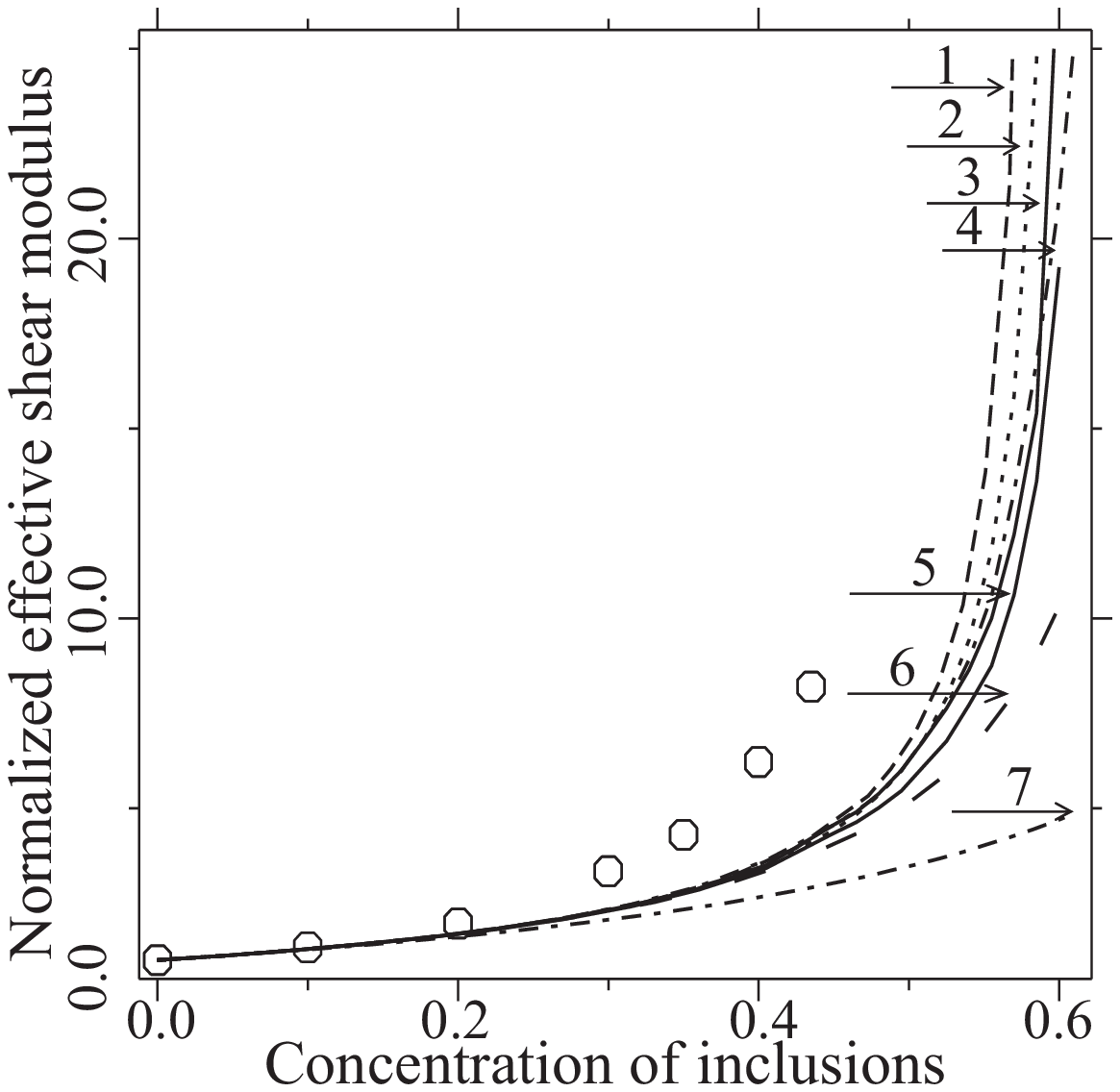,width=12.cm}\\
\vspace{-80.mm}\tenrm \baselineskip=8pt

{\rm Fig. 2}. $\mu^*(c)/\mu^{(0)}$ vs $c$ estimated by the MEFM with RDF simulated by
Willis approximation (1), SCRM's approximation (5.5) (2), CRM (3), SCRM (4), PY (5), well-stirred approximation (6), and by the MTM (7); $\circ$ - experimental data by Kreger [63]
}
\vspace{2mm}

\noindent RDF $g(r,c)$ ($c\equiv 0.5$) according to Eq. (17I).
As can be seen, the curve 2 provides a reasonably accurate approximation of the curve 4 corresponding to the SCRM. The effective moduli $\mu^*$ {\color {black}as the functions of $c$ ($\mu^*\sim c$)} are very
sensitive with respect to the RDF; for example, the use of well-stirred approximation (37) leads to reduction of the predicted $\mu^*(c)$ at $c=0.6$ at least by the factor 2 or 3 with respect to the RDFs estimated by Eq. (39) and by CRM, and SCRM. Neglecting by the binary interaction of inclusions (27) {\color{black}leads to} subsequent diminution of $\mu^*(0.6)$ in two times.
All RDFs taking a neighboring order into account lead
to infinite values of $\mu^*(c)$ for large values of $c$, however the limiting values $c\simeq 0.61$ and $c\simeq 0.62$ corresponding the CRM and SCRM, respectively, are better compatible with an {\color{black}experimental} value $c_m=0.637$ than the limiting assessments $c=0.58$ for the Willis approximation (39).

\bigskip
\noindent{\textit{\textbf{ 4.2 Effective moduli and stress concentrator factors of composites with bimodal packs}}
\medskip

We estimate in Fig. 3
 the relative effective shear moduli $\mu^*/\mu^{(0)}$ of composites with incompressible matrix containing rigid spherical inclusions with the RDFs (38)  as well as with the RDF simulated by the SCRM (I); $\lambda_{21}\equiv a^{(2)}/a^{(1)}=0.313$ and $\xi_{21}\equiv c^{(2)}/c^{(1)}=0.333$.

As can be seen, the values $\mu^*/\mu^{(0)}\sim c$ estimated by the MEFM for the RDFs corresponding to the SCRM's simulation correlates better with experimental data by Chong, Christiansen and Baer (1971) than
analogous estimations for the RDF (38). Although the effective behavior of the composite is traditionally the main focus of micromechanics, it is also essential to supply insight into the statistical description of the local stresses in each phase and at interphase.
Estimation of these local fields are extremely useful
for understanding the evolution of nonlinear phenomena such as plasticity, creep, and damage (see, e.g, Ravichandran and Liu, 1995; Tan {\it et al.}, 2005, 2007).
We present Eq. (24) in the form
\BB
\lle\bfsi\rle_k=\bfB{\bfD}_k\lle\bfsi\rle, 
\EE
 where the tensor $\bfD_k$ ($k=1,2$) (called effective stress concentrator factor) has a simple physical meaning of the action of surrounding inclusions on the separate one $\lle\overline{\bfsi}\rle_k=\bfD_k\lle\bfsi\rle$. Equation (40) makes it possible to
estimate the ensemble average of the stress in the vicinity of the inhomogeneities
near the point $\bfx\in v_i\subset v^{(k)}$ with the unit outward normal vector $\bfn\perp\partial v_i$ ($k=1,2$)
\BB
\lle\bfsi^{-}(\bfn)\rle_{\bf x}=\bfB(\bfn)\bfB{\bfD}_k\lle\bfsi\rle, 
\EE
where the tensor $\bfB(\bfn)$ depends only on the elastic properties of the
contacting materials and by the direction of the normal $\bfn$ (see for details Buryachenko, 2007a).
Thus, the stress concentrators factors inside (40) and outside (41) inclusions depend on the linear functions of both the elastic solutions $\bfB$, $\bfB(\bfn)$ for a single inclusion inside infinite matrix and the concentrator factor $\bfD_k$ defined intrinsically by a micromechanical model. Because of this, we will only analyze a dependence $\bfD_k\sim c$ ($k=1,2$).

 As can be seen in Fig. 4, the {\color{black} components $D_{k|1111}$ of the stress concentrator factors $\bfD_k$ (40)} for both the RDF (38) and SCRM's
RDF at the large ($k=1$) and small ($k=2$) inclusions can be distinguished one from another by tens {\color{black}of} percent that has a significant practical implementation.
Indeed,
for periodic structures, Zhong and Knaus (2000) numerically demonstrated that for a large difference in particle sizes (such as a bimodal distribution), damage
occurs at interfaces between the big particles and matrix, while only limited or no
damage occurs at interfaces around small ones (that is also experimentally confirmed, see Draughn, 1981; Lewandowski, Liu and Hunt, 1989). While these effects of particle
interaction and size variation are smoothed out in a large ensemble of particles that are exhibited by the effective elastic moduli,
it is foreseeable that they are an important factor in a failure process such as local {\color{black}debonding} when the particle size {\color{black}variation}
leads to a yield-likely response characterized by a nonmonotonic load deformation trace.
 Such a characteristic may, ultimately, be associated with inhomogeneous deformation
such as occurs in shear bands or other strain concentrations, for example. The effect of increased stress concentrator factors at the big particles for random structure bimodal structures pioneered in the current paper is critical for any nonlinear localized
phenomena such as onset of yielding, failure initiation, and others. Of prime importance is
that the mentioned size distribution effect was discovered for linear elastic composites with ideal contact between the matrix and particles (analysis of non-ideal contact, see e.g. Tan {\it et al.}, 2005, 2007, is beyond the scope of the current publication).

\vspace{-17mm}
\noindent \parbox{7.8cm}{
\hspace{-25mm} \psfig{figure=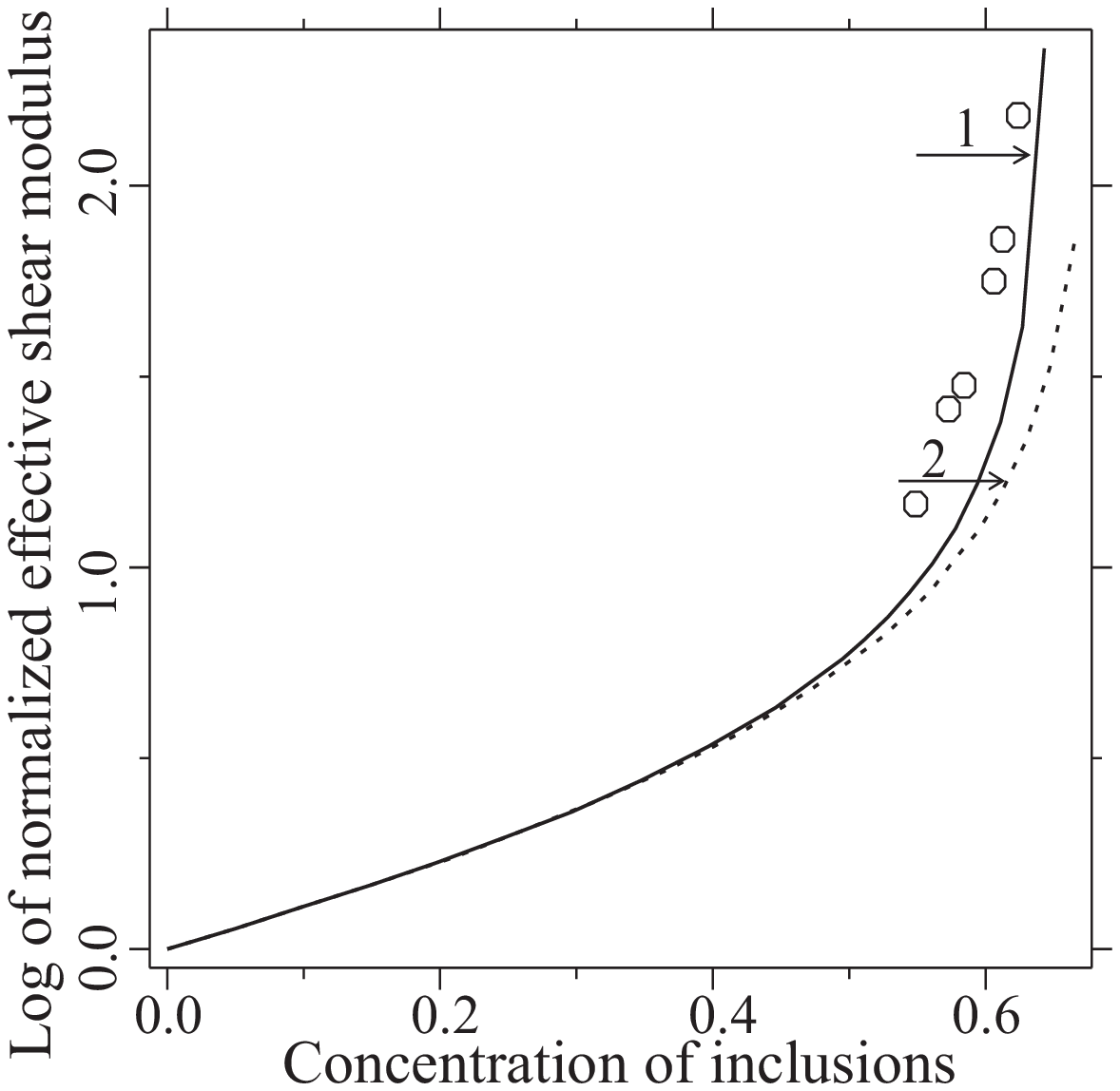, width=12.cm}\\
\vspace{-80mm} \tenrm \baselineskip=8pt

 {\rm Figure 3}:
 ${\rm Log}[\mu^*(c)/\mu^{(0)}]$ vs $c$ estimated by the MEFM with RDFs simulated by the SCRM (curve 1), and Eq. (30)
 (curve 2); $\circ$ - experimental data by Chong, Christiansen and Baer (1971)
 }
 \ \ \noindent \parbox{7.8cm}{
\vspace{-7mm} \hspace{-20mm}
\psfig{figure=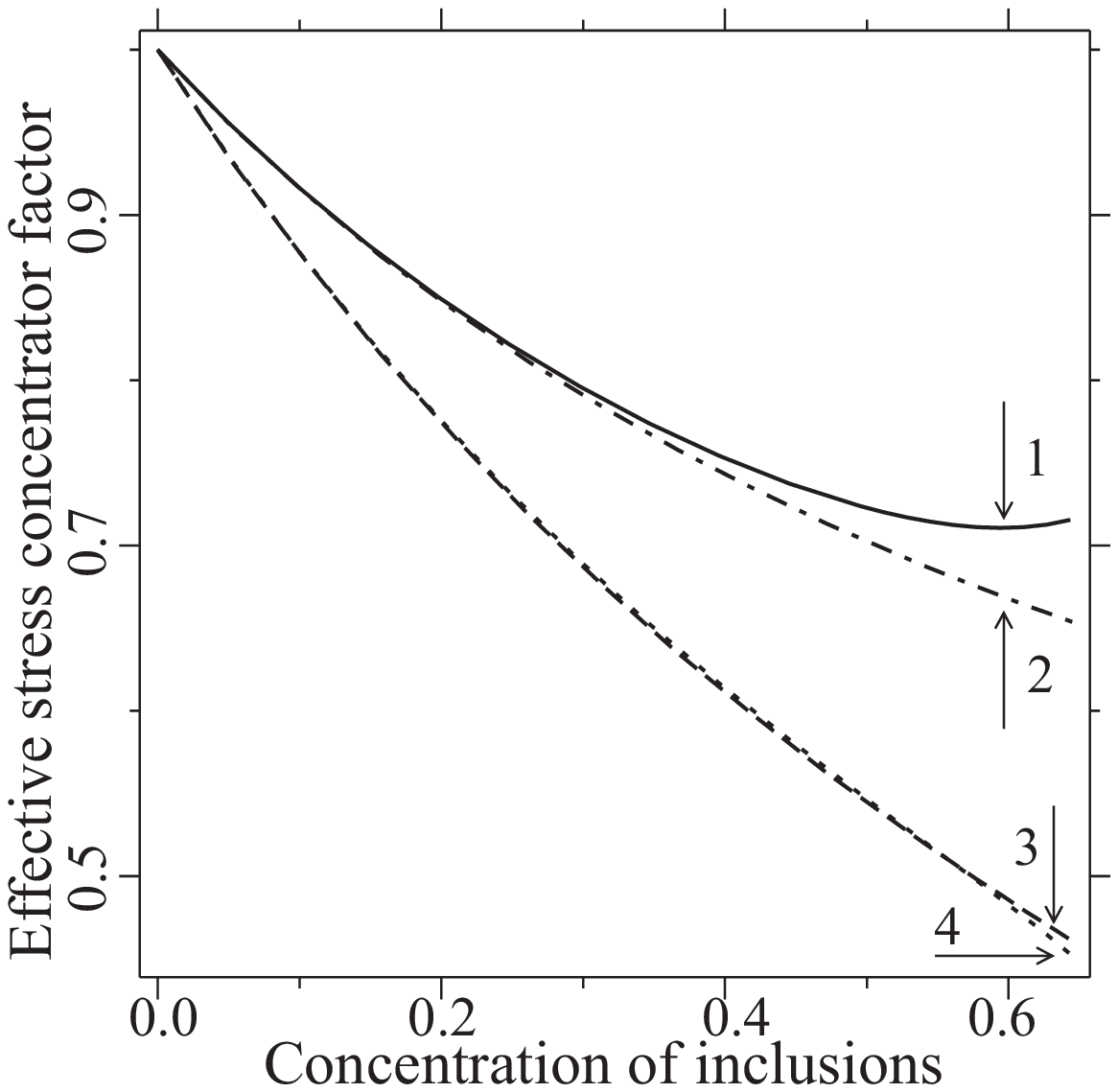,width=12.cm}\\
\vspace{-80.mm}\tenrm \baselineskip=8pt

{\rm Fig. 4}. $D_{k|1111}$ vs $c$ estimated
 for the RDFs simulated by the SCRM (curves 1 and 4), and Eq. (26) (curves 2 and 3) at the big inclusions (1,2) and small ones (3,4) }
\vspace{2mm}

\medskip
\noindent{\bf 5 Conclusion}
\medskip

\noindent{\textit{\textbf{ 5.1 Some critical comments}}
\medskip

It does not need to be emphasized that popular methods using the second-order probability function $S_{ij}(r)$ ($i,j=0,1,\ldots,N$) (1I) (see for references Buryachenko, 2007a; Lee, Gillman and Matou\v{s}, 2011) are invariant with respect to both $S_{ij}(r)$ [at least for statistically isotropic structures, see, e.g., Eqs (36) and (37)] and the size distribution of inclusions. Because of this, numerous estimations of $S_{ij}(r)$ (see e.g. Kumar, Matou\v{s} and Geubelle, 2008; Lee {\it et al.}, 2009; Lee, Gillman and Matou\v{s}, 2011; Matou\v{s} and Geubelle, 2006; Liu {\it et al.}, 2012) have only a marginal practical use from the perspective of view of micromechanics (see Fig. 2 and the comments after Eq. (37)). Moreover, a potential usefulness of $S_{ij}(r)$'s estimations is reduced by a simplified assumption of a two-phase approximation $(i,j=0,1$) (see. e.g., Jackson, Hooks and Buckmaster, 2011) of the 2-point probability density $S_{ij}(r)$ even for multimodal distribution ($N>1$) of particle sizes.
$S_{ij}(r)$ is just one of a few possible statistical descriptors which is significantly less sensitive to the structure microtopology than the RDF that provides a smoothness of $S_{ij}(r)$ (in opposite to the RDF's estimations which smoothness can be only achieved by additional sophisticated efforts, e.g., both the parallelization of simulations and the shaking procedure, see (I)). Smoothness $S_{ij}(r)$ and its poor  sensitivity in comparison with RDF in analysis of random structures provide  a possibility for concealment of disadvantages in this simulation. Moreover, coincidence of the $S_{ij}(r)$ in both the real sample and the reconstructed structures generated by the protocol dependent algorithms (see, e.g.,  Kumar, Matou\v{s} and Geubelle, 2008; Lee {\it et al.}, 2009) does not assure a coincidence of other more important statistical descriptors (e.g. RDFs). Because of this, comparison of RDFs in both the structure generated by protocol independent algorithm
  (see (I)) and reconstructed structure can be considered as a validation of the protocol dependent algorithm used for configuration reconstruction. Furthermore, geometrical statistical equivalence of the real and reconstructed packs is necessary but not sufficient condition for micromechanical statistical equivalence. In fact, Lee {\it et al.} (2009), and Lee, Gillman and Matou\v{s} (2011)
 implicitly introduced (with an intuitive level of rigor)
 a new concept of Representative Volume Element (RVE) for geometrical statistical equivalence. This concept is not well defined and depends on the statistical descriptors used. However, the concept RVE is well known in micromechanics for micromechanical statistical equivalence
 (see for references Buryachenko, 2007a) which is defined not only by $S_{ij}(\bfx_i-\bfx_j)$ (or $\varphi(v_j,\bfx_j|;v_i,\bfx_i)$) but also by an elastic particle interaction described by $\varphi(v_j,\bfx_j|;v_i,\bfx_i)$  rather than by $S_{ij}(\bfx_i-\bfx_j)$.

  It should be mentioned a growing in importance of advanced experimental techniques (such as X-ray tomography and electron microscopy) in analysis of random structures of HSP. In particular, the papers referred above (see also for references (I)) demonstrate a large opportunities of statistical descriptions of real structures by $S_{ij}(\bfx_i-\bfx_j)$. However, a practical using of this statistical descriptors $S_{ij}(\bfx_i-\bfx_j)$ estimated in real HSP is even more important.  In this sense it is expected a significant impact on the scientific society of the paper Lee, Gillman and Matou\v{s} (2011) where the authors proposed a direct substitution of $S_{ij}(\bfx_i-\bfx_j)$ estimated in the real packs into the corresponding formulae of analytical micromechanics (29) where according to the hypothesis {\bf H3} $S_{ij}(\bfx_i-\bfx_j)$ should have the ellipsoidal symmetry (33).

  A popular explanation of acceptance of this "ellipsoidal symmetry" hypothesis (33) is that this hypothesis
just simplifies Eq. (32) reducing this equation to Eq. (35) which does not contain the integrals. In a similar
manner, a destination of the assumption of the ellipsoidal shape of the excluded volume $v^0_i$
 in the hypothesis
{\bf H3} is that this hypothesis just simplifies Eq. (35) by the use of the analytical known tensor $Q(v^0_i)$ (expressed through the Eshelby tensor $\bfS(v^0_i)$ (21)) which is exploited instead of a general tensor $Q(v^0_i)(\bfx)$ found numerically for an arbitrary shape of $v_i^0$
(see e.g., Subsection 4.7.4 in Buryachenko, 2007a). However, both mentioned assumptions of the hypothesis {\bf H3} have a
fundamental conceptual sense rather than only an analytical solution of some particular problem. Exploiting the
Eshelby tensor concept in Eq. (4.18) (and in the MEFM) is based on the ellipsoidal shape of the correlation hole
$v^0_i$ rather than on the inclusion shape $v_i$.
An abandonment of either the assumption of the ${v^0_i}$'s ellipsoidal shape
or "ellipsoidal symmetry" hypothesis (32) necessarily leads to the inhomogeneity of the effective field $\overline{\bfsi}_i(\bfy)$
$(\bfy\in v_i)$ acting on the inclusion $v_i$ that is prohibited for the classic background of micromechanics based on Eq. (9) (see for details Subsection 5.3 and Buryachenko, 2010a, 2010b, 2010c, 2012).

Thus, correct performing of micromechanical analysis in Section 3 uniquely determines the form of statistical descriptors either
$\varphi(v_j,\bfx_j|;v_i,\bfx_i)$  or $S_{ij}(\bfx_i-\bfx_j)$ estimated from the real packs.
At first, it is necessary to check that the real pack has a statistically anisotropic structure rather than a functional graded configuration (as in Lee, Gillman and Matou\v{s}, 2011) requiring a different level of micromechanical modeling (see for details Buryachenko, 2007a).
For statistically anisotropic packs, the mentioned descriptors must be approximated by the appropriate functions with ellipsoidal symmetry (33), otherwise the micromechanical models corresponding to classic background of micromechanics (9) can not be used (although the new background of micromechanics considered in Subsection 5.3 makes it possible to exploit any descriptors). In any way, $\varphi(v_j,\bfx_j|;v_i,\bfx_i)$  is preferred over $S_{ij}(\bfx_i-\bfx_j)$ due to better sensitivity to a concrete microstructure
(either statistically isotropic, statistically anisotropic, or functionally graded) as well as to exploiting precisely $\varphi(v_j,\bfx_j|;v_i,\bfx_i)$ (rather than $S_{ij}(\bfx_i-\bfx_j)$) in the advanced micromechanical methods (see Fig. 2 and Buryachenko, 2007a, 2011).

\medskip
\noindent{\textit{\textbf{ 5.2 Some prospective problems}}
\medskip

{\color{black} The slightly modified version of the MEFM proposed provides the calculation with reasonable accuracy of the effective elastic moduli and statistical average of
stresses in the components for the composites with bimodal distribution of spherical particles. A straightforward generalization of this model is possible to the multimodal particle distribution. It also allows the model to simultaneously account for a plethora of constituents in the system - for example, metal powder, short fibers, and voids can be included.

However, the most promising result of the approach proposed is sensitivity of estimations of both the effective elastic moduli and stress concentrator factors on the size ratios of particles that is especially critical for the HSP with extremely high volume fractions ($>\!0.9$). A similar effect is expected to be detected for other linear problems such as e.g. thermoconductivity, thermoelasticity, viscoelasticity (Banerjee and Adams, 2004) which are concerned with a number of propellant-related issues, including the burning rate and the thermomechanical loading of the HSPs.  Intensively investigated micromechanical problem with imperfect interface between the matrix and particles are solved by substitution into the known cohesive zone model the statistical averages of the local stresses at the oxidizer particles estimated by the Mori-Tanaka method (see for references Tan {\it et al.}, 2005, 2007). These estimations (depending only on the volume fraction $c$ rather than on the RDFs $g_{ij}$ and  the size ratios $\lambda_{ij}$ of particles) can be significantly refined by the proposed version of the MEFM.

Evolution of nonlinear phenomena such as  viscoplasticity, creep, and damage are a new challenge for designers of rockets. When one tries to estimate the equivalent
stress in the strength theories as well as in nonlinear creep theory, or when the yield function in plasticity theory is considered, squares of the first invariant or
the second invariant of the deviator of local stresses are frequently used. For unimodal distribution of particles, estimations of the required second statistical moments of local stresses can be performed by both the perturbation method and the method of integral equations (see for details Chapter 13 in Buryachenko, 2007a). Generalization of the mentioned methods to the case of the multimodal distribution of particles (which is inherent in the HSP) is  straightforward. It opens a new avenue of questions on some specific nonlinear problems such as
ignition, hot spot analysis, and detonation. So Field {\it et al.} (1992),
Massoni {\it et al.}, 1999; Tan {\it et al.} (2005), Dienes, Zuo, and Kershner (2006), Grinfeld (2009) indicated the following
mechanisms of energy concentrations in locations called ``hot-spot" causing localized melting, ignition, and fast burn of the reactive material: 1) Adiabatic compression of cavity gases; 2) Heating of solid adjacent to collapsing cavity; 3) Viscous heating of binder between grains; 4) Friction between impacting surfaces; 5) Localized adiabatic shear; 6) Heating by frictional sliding between the faces of a closed cracks and debondings; 7) Debonding of particles and matrix. All of these problems were partially solved for HSPs and explosives by simplified micromechanical models such as, e.g.,
the dilute approximation ($\lle\overline {\bfsi}\rle_i=\lle\bfsi\rle$), the effective medium approximation considering each defect inside the effective medium ($\bfL^{(0)}=\bfL^*$), and the mixture role of damage mechanics. It can also be mentioned that substitution of estimations of both the first and second statistical moments of local stresses performed by the proposed approach (see Sections 3 and 4) into the mentioned nonlinear problems is also straightforward.

However, more detailed consideration of particular problems mentioned is beyond the scope of the current paper and will be considered in subsequent publications.
More rich in content is a discussion of  the main hypotheses  as well as the limitations of
the proposed estimations and their possible generalizations for solutions of some prospective problems.}

\medskip
\noindent{\textit{\textbf{5.3 Generalization of some hypotheses}}
\medskip

The assumption {\bf H1} of homogeneity of $\overline {\bfsi}({\bf x}),\ ({\bf x}\in v_i)$
is a classical hypothesis of micromechanics (see for the earliest references
Lax, 1952) and was accepted in order to make it easier to solve the problems for both one (19) and two (22) particles.
Buryachenko (2007a) demonstrates that the MEFM includes plurality of popular methods based on the hypothesis {\bf H1}.
The accuracy of this hypothesis was estimated for two spherical inclusions in the infinite matrix
as well as for the periodic structure composites (see for references Buryachenko, 2007a).

{\color{black}However, the hypothesis {\bf H1)} is merely a zero-order approximation of binary interacting inclusions that
results in a significant shortcoming of the MEFM. This substantial
obstacle can be overcame in the
upgraded version of the MEFM proposed by Buryachenko (2007b) in light of the generalized schemes based
 on the numerical solution (some sort of the building blocks) of
the problem for both one and  two inclusions in the infinite media, subjected to the effective field evaluated from
forthcoming self-consistent estimations. It is possible to get a numerical solution in the form
of a table in which each line among a few hundred lines corresponds to an accurate solution for
two inclusions with concrete coordinates in an infinite medium subjected to the unit loading.
No restrictions are imposed on the microtopology of the microstructure and the shape
of inclusions as well as on the inhomogeneity of stress field inside the inclusions.
However, the main computational advantage of the generalized MEFM
 lies in the fact that such fundamental notions of micromechanics as a Green function and Eshelby
tensor are not used, and we can analyze any anisotropy of constituents
(including the matrix) as well as any shape and any composite structure of inclusions.
 The known numerical methods such as FEA, VIE, BIE, and the multi,
 multipole expansion method (see e.g., Hatch and Davis, 2006; Buryachenko,  Kushch, and Roy, 2007), and complex potential method
 (Buryachenko and Kushch, 2006) which can be used for construction of the building blocks mentioned,
have a series of advantages and disadvantages. It is crucial for the analyst to be aware of their range of applications.
In particular, for the high volume fraction of particles $c$, phase arrangements will involve frequent direct contacts between neighboring particles
and may even lead to a local stress singularities between the contacting rigid particles. However, these strong inhomogeneity of stress distributions can only be detected and analyzed
by advanced accurate numerical methods (see, e.g.,  Hatch and Davis, 2006; Buryachenko and Kushch, 2006) while the analytical method (22) used in the current paper takes into account
only the stresses averaged over the volumes of neighboring particles. The accuracy of the method (22) estimated in Fig. 4.4 in Buryachenko (2007a) eliminates the possible difficulties
which may arise when direct contact between neighboring particles occurs frequently.}

Fundamentally more general approach is based on a new general integral equation of micromechanics
proposed by Buryachenko (2010a, 2010b)
\BB
\bfsi ({\bf x})=
\langle \bfsi\rangle
+\int \Big[ {\bf \Gamma} (\bfx-\bfy) \bfeta({\bf y})
-\langle {\bf \Gamma} (\bfx-\bfy)\bfeta\rangle ({\bf y})\Big] d{\bf y},
\EE
Equation (34) was obtained without any auxiliary
assumptions such as, e.g., the version of the {\bf H1}: $\lle\bfGa(\bfx-\bfy)\bfeta(\bfy)\rle(\bfy)=
\bfGa(\bfx-\bfy)\lle \bfeta\rle(\bfy)$ (hypothesis {\bf H1b}, p.
253 in Buryachenko, 2007a) implicitly exploited in the known centering methods and reducing Eq. (42) for statistically homogeneous
media subjected to the homogeneous boundary conditions to the known Eq. (9)
which goes back to Lord Rayleigh (1892). Buryachenko (2010a, 2010b) demonstrated  that Eq. (9), erroneously recognized
as an exact one after the proofs by Shermergor (1977), is correct only after the additional asymptotic assumption H1b.
What seems to be only a formal trick is in reality a new background of micromechanics
allowing us to abandon the
central concept of classical micromechanics, such as effective field
hypothesis {\bf H1}. Then the hypothesis {\bf H2} taking into account the
binary interaction of inclusions leads to the inhomogeneity of both the statistical
average stresses inside the inclusions and the effective fields (violation of
the effective field hypothesis {\bf H1}) even for statistically homogeneous composites subjected to the
homogeneous remote loading and containing homogeneous ellipsoidal inclusions. It is
expected significant improvement of accuracy of statistical averages of local stress concentrator
factors (with a possible change of the sign of predicted local stresses) in the
approach by Buryachenko (2011) (only applied for $2D$ case) with respect to a
classical one presented in this paper while their averaged values $\lle\bfsi\rle_i$
(and, therefore, the effective moduli $\bfL^*$) are only slightly sensitive to the
random fiber arrangement described by the RDF.

Here a question of verification
of the approach proposed appears. Unfortunately, validating the numerical results
expected is currently beyond the capabilities of standard experimental techniques
based on the measurement of effective elastic moduli $\bf L^*$ because
the difference of $\bfL^*$ estimated by the new Buryachenko (2011) and Buryachenko (2007a)
approaches is expected negligible (at least for the moderate volume fraction $c$).
However, the advantage of the method proposed can be seen in the fact that the emphasis in materials modeling shifted from explaining experimentally observed properties of materials (such as $\bfL^*$) to predicting not-yet-measured properties (such as the local stresses).
Therefore, there is an increasing need to provide very detailed experimental
information about the local stress concentrator factors of numerically investigated
materials at the microscale.
New volumetric digital image correlation (VDIC) relying on the X-ray tomographic
imaging of naturally occurring material texture within samples requires the
ability to reproduce the position of image points with ``high precision" because the
determination of relative deformations requires two images for comparison purposes to
extract strain measures. Full-field measurements of strains by VDIC (and by other
imaging tools) are potentially well suited to analyze the specific mechanical
properties of composite materials (see, e.g., Gr\'ediac, 2004; Sutton, Orteu and Schreier, 2009).
However, the mentioned opportunities of experimental methods are not yet realized at the level required for verification of the new approach mentioned above.

Of course, all particular problems mentioned in Subsections 5.1 and 5.2 can be solved in the framework of the new background of micromechanics indicated in the current subsection 5.3 (see also Buryachenko, 2011, 2012).


\medskip
\noindent{\bf Acknowledgment:}
This work was partially funded by IllinoisRocstar LLC and Micromechanics \& Composites LLC. Both the helpful comments of reviewers and their encouraging recommendations are gratefully acknowledged.

\medskip
\noindent{\bf References}
\medskip

{\baselineskip=9pt
\parskip=1pt

\hangindent=0.4cm\hangafter=1\noindent
{\bf Absi, E.; Prager, W. } (1975): A comparison of equivalencies and
finite element methods. {\it  Comput. Meth. Appl. Mech. Eng.},
 vol.{ 6}, pp. 15--99

 \hangindent=0.4cm\hangafter=1\noindent{\bf
Annapragada, S.R.; Sun, D.; Garimella, S.V. } (2007):
Prediction of effective thermo-mechanical properties of particulate composites.
{\it  Computational Materials Science},  vol.{ 40}, pp. 255--266

 \hangindent=0.4cm\hangafter=1\noindent{\bf
 Arefinia, R.; Shojaei, A.} (2006)
On the viscosity of composite suspensions of aluminum and ammonium
perchlorate particles dispersed in hydroxyl terminated
polybutadiene - New empirical model.
{\it J. Colloid and Interface Sci.}, vol.  299, 962-–971

\hangindent=0.4cm\hangafter=1\noindent{\bf
Banerjee, B.; Adams, D. O. } (2004):
On predicting the effective elastic properties of polymer
bonded explosives using the recursive cell method
{\it  Int. J. Solids Structures},  vol.{ 41}, pp. 481--509

\hangindent=0.4cm\hangafter=1\noindent{\bf
Belytschko, T.;  Black, T. } (1999):  Elastic crack growth in finite elements with minimal remeshing.
{\it Int. J. Numer. Methods Engrg.},  vol.{ 45}, pp. 601--620

\hangindent=0.4cm\hangafter=1\noindent{\bf
 {Benveniste, Y.} } (1987):  A new approach to application of
Mori-Tanaka's theory in composite materials. {\it  Mech. Mater.}
 vol.{ 6}, pp. 147--157

\hangindent=0.4cm\hangafter=1\noindent
{\bf  B\"ohm, H. J.; Han, W.; {\color{black}Eckschlager}, A.} (2006):
Multi-inclusion unit cell studies of reinforcement stresses and particle
failure in discontinuously reinforced ductile matrix composites.
{\it CMES: Computer Modeling in Engineering \& Sciences}, vol. 5, pp. 5--20

\hangindent=0.4cm\hangafter=1\noindent{\bf
Borb\'ely, A.; Kenesei, P.; Biermann, H. } (2006):
Estimation of the effective properties of particle-reinforced
metal matrix composites from microtomographic reconstructions.
{\it  Acta Materialia},  vol.{ 54}, pp. 2735--2744

\hangindent=0.4cm\hangafter=1\noindent{\bf
Buryachenko, V. A. } (2007a):  {\it  Micromechanics of Heterogeneous Materials}. Springer, NY.

\hangindent=0.4cm\hangafter=1\noindent{\color{black}{\bf
Buryachenko, V.A.} (2007b): Generalization of the multiparticle effective field method
in static of random structure matrix composites. {\it Acta Mechan.}, vol. {188}, pp. 167-208}

\hangindent=0.4cm\hangafter=1\noindent{\color{black}{\bf
Buryachenko V. A.} (2010a): On some background of micromechanics of random structure matrix composites.
{\it Int. J. Pure and Applied Mathematics}., vol. {59}, pp.  163--179}

\hangindent=0.4cm\hangafter=1\noindent{\color{black}{\bf
Buryachenko, V.A.} (2010b): On the
thermo-elastostatics of heterogeneous materials. I. General
integral equation. {\it Acta Mech}., vol.  {213}, pp. 359--374}

\hangindent=0.4cm\hangafter=1\noindent{\color{black}{\bf
Buryachenko, V.A.} (2010c): On the
thermo-elastostatics of heterogeneous materials.
II. Analyze and generalization of some basic hypotheses and propositions.
{\it Acta Mech}., vol.  {213}, pp. 375--398}

\hangindent=0.4cm\hangafter=1\noindent{\bf
Buryachenko, V. A. } (2011):  Inhomogeneity of the first and second statistical moments of stresses inside the heterogeneities of random structure matrix composites. {\it  Int. J. Solids Struct.},  vol.{ 48}, pp. 1665--1687

\hangindent=0.4cm\hangafter=1\noindent{\bf
Buryachenko, V.} (2012): General integral equations of micromechanics of composite materials with imperfectly bonded interfaces. (Submitted).

\hangindent=0.4cm\hangafter=1\noindent{\bf
Buryachenko, V.;  Jackson, T.;   Amadio, G.} (referred to as (I) in the text) (2012):
Modeling of random bimodal structures of composites
(application to solid propellant):
 I. Simulation of random packs. {\it Computer Modeling in Engineering \& Sciences}. (In Press).

 \hangindent=0.4cm\hangafter=1\noindent{\bf
  Buryachenko V. A.;  Kushch V. I.} (2006): Effective transverse elastic moduli of composites at
  non-dilute concentration  of a random field of aligned fibers.  {\it ZAMP}, vol. 57, pp. 491--505

\hangindent=0.4cm\hangafter=1\noindent{\bf Buryachenko, V. A.; Kushch, V. I.; Roy. A.} (2007):
Effective thermoelastic properties of random
structure composites reinforced by the clusters of deterministic structure (application to clay
nanocomposites). {\it  Acta Mechanica}, vol. 192, pp. 135–-167

\hangindent=0.4cm\hangafter=1\noindent{\bf
Buryachenko, V. A.; Pagano, N. J.; Kim, R. Y.; Spowart, J. E. } (2003):  Quantitative description of random microstructures of composites and their effective elastic moduli. {\it  Int. J. Solids Struct.},
 vol.{ 40}, pp. 47--72

\hangindent=0.4cm\hangafter=1\noindent{\bf
Chen, H.-S.; Acrivos, A. } (1978):  The solution of the equations of linear elasticity for an infinite region,
containing two spherical inclusions. {\it  Int. J. Solids Structures},  vol.{ 14}, pp. 331--348

\hangindent=0.4cm\hangafter=1\noindent{\bf
Chen, X.; Papathanasiou, T. } (2004):  Interface stress distributions in transversely loaded continuous
fiber composites: parallel computation in multi-fiber RVEs
using the boundary element method.
{\it  Composites Science and Technology},  vol.{ 64}, pp. 1101--1114

\hangindent=0.4cm\hangafter=1\noindent{\bf
Chong, J.S.; Christiansen, E.B.; Baer AD. } (1971):  Rheology of concentrated suspensions.
 {\it  J. Applied Polymer Science},  vol.{ 15}, pp. 2007--2021

\hangindent=0.4cm\hangafter=1\noindent{\bf
Clements, B.E.;  Mas, E.M.} (2004):
A theory for plastic-bonded materials with a bimodal
size distribution of filler particles
{\it Modelling Simul. Mater. Sci. Eng.}, vol.  12, pp.   407–-421

\hangindent=0.4cm\hangafter=1\noindent{\bf
 D'Haene, P.; Mewis, J. } (1994):  Rheological characterization of bimodal colloidal dispersions.
{\it  Rheol Acta},  vol.{ 33}, pp. 165--174

\hangindent=0.4cm\hangafter=1\noindent{\bf
Dienes, J.K.; Zuo Q.H.; Kershner, J.D.} (2006):
Impact initiation of explosives and propellants via statistical crack mechanics.
{\it J.  Mech. Physics  Solids}, vol: 54, pp. 1237--1275

\hangindent=0.4cm\hangafter=1\noindent{\bf
Draughn, R.A. } (1981):
Effects of temperature on mechanical properties of
composite dental restorative materials. {\it  J. Biomedical Mater. Research}
 vol.{ 15}, pp. 489--495

\hangindent=0.4cm\hangafter=1\noindent{\bf
Du, C.; Ying, Z.; Jiang, S. } (2010):
Extended finite element method and its application in heterogeneous materials with inclusions.
{\it  IOP Conf. Series: Materials Science and Engineering.},  vol.{ 10}, 012083

\hangindent=0.4cm\hangafter=1\noindent{\bf
Enikolopyan, N.S.; Fridman, M. L; Stalnova, I.O.; Popov, V.L. } (1990):  Filled polymers:
mechanical properties and processability. {\it  Adv. Polym. Sci.},  vol.{ 96}, pp. 1--67

\hangindent=0.4cm\hangafter=1\noindent{\bf
  Eshelby J. D.}  (1961):  Elastic inclusion and inhomogeneities In: {I. N. Sneddon}, {R. Hill} (eds.)
{\it Progress in Solid Mechanics}, vol  2,
pp.89--140.
 North-Holland, Amsterdam.

\hangindent=0.4cm\hangafter=1\noindent{\bf
Farris, R.J. } (1968):  Prediction of the viscosity of multimodal suspensions
from unimodal viscosity data. {\it  Trans. Society of Rheology},  vol.{ 12}, pp. 281--301

\hangindent=0.4cm\hangafter=1\noindent{\color{black}{\bf
Field, J.E.; Bourne, N.K.; Palmer, S.J.P.; Walley, S.M.} (1992): Hot-spot ignition mechanisms for explosives and propellants.
{\it Philos. Trans. R. Soc.}, vol:  A339, pp. 269--283}

\hangindent=0.4cm\hangafter=1\noindent{\bf
Franciosi, P.; Lebail, H. } (2004):  Anisotropy features of phase and particle spatial pair distributions in various matrix/inclusions structures. {\it  Acta Mater},  vol.{ 52}, 3161--3172

\hangindent=0.4cm\hangafter=1\noindent{\bf
Fries, T.P.; Belytschko, T. } (2010):  The extended/generalized finite element method: An overview of the method and its applications.
{\it  Int. J. Numerical Methods Engineering}, {\tenbf
84}, pp. 253--304

\hangindent=0.4cm\hangafter=1\noindent
{\color{black}{\bf Garishin, O.C.} (1997):
Physical discretezation and its application in the structural model of composite. In: Moshev, V.V., Svistkov, A.L., Garishin, D.C., Evlampieva, S.E., Rogovoy, A.A., Kovrov, V.N., Komar, L.A., Golotina, L.A., Kozhevnikov A.L.
{\it Structural mechanisms of the mechanical properties of particulate polymer composites}, Ural Branch of RAS, Perm, pp. 80--168 (In Russian)}

 \hangindent=0.4cm\hangafter=1\noindent{\bf
Garishin, O.C.; Moshev, V.V. } (2002):  Damage model of elastic rubber particulate composites.
{\it  Theoretical Applied Fracture Mechanics}.
 vol.{ 38}, pp. 63--69

\hangindent=0.4cm\hangafter=1\noindent{\bf
Ghosh, S. } (2011):
{\it  Micromechanical Analysis and Multi-Scale Modeling Using the Voronoi Cell Finite Element Method (Computational Mechanics and Applied Analysis).}
CRC Press, Boca Raton

 \hangindent=0.4cm\hangafter=1\noindent{\bf
Gr\'ediac, M. } (2004):  The use of full-field measurement methods in composite material characterization: interest and limitations. {\it  Composites},  vol.{ A35}, pp. 751--761

 \hangindent=0.4cm\hangafter=1\noindent
 {\color{black}{\bf
 Grinfeld, M.A.} (2009): Voids as triggers of hot spots in solids.
  {\it Shock Compression of Condensed Matter}, edited by M. L. Elert, W. T. Buttler, M. D. Furnish, W. W. Anderson, and W. G. Proud. pp. 201--204}

 \hangindent=0.4cm\hangafter=1\noindent
 {\color{black}{\bf
  Gusev, A.; Lusti, H.R.; Hine, P.J.} (2002): Stiffness and thermal expansion of short fiber
  composites with fully aligned fibers. {\it Adv. Eng. Mater.}, vol. 4, pp. 927-–931}

 \hangindent=0.4cm\hangafter=1\noindent
 {\bf
 H\"afner, S.; Eckardt, S.; Luther,T.; K\"onke, C. } (2006):
Mesoscale modeling of concrete: Geometry and numerics.
{\it  Computers Structures},  vol.{ 84}, pp. 450--461

\hangindent=0.4cm\hangafter=1\noindent
{\bf
Hashin, Z.; Shtrikman, S. } (1963):  A variational approach to the theory of the behavior
of multiphase materials. {\it  J. Mech. Phys. Solids.},  vol.{ 11}, pp. 127--140

\hangindent=0.4cm\hangafter=1\noindent
{\color{black}{\bf
Hatch, R.L.; Davis, I. L.} (2006):
Mechanical properties for an arbitrary arrangement of rigid spherical
particles embedded in an elastic matrix. {\it (Preprint).} ATK Launch Systems
Group,  Brigham City, UT}

\hangindent=0.4cm\hangafter=1\noindent
{\bf
He, D.; Ekere, N .N. } (2001):
Structure simulation of concentrated suspensions of hard spherical particles.
{\it  AIChE Journal}  vol.{ 47}, pp. 53--59

\hangindent=0.4cm\hangafter=1\noindent
{\bf
{Hill, R.} } (1965):  A self-consistent mechanics of composite
materials. {\it  J. Mech. Phys. Solids}  vol.{ 13}, pp. 212--222

\hangindent=0.4cm\hangafter=1\noindent
{\color{black}{\bf
Hiriyur, B.;  Waisman, H.; Deodatis, G.} (2011):
Uncertainty quantification in homogenization of heterogeneous
microstructures modeled by XFEM.
{\it Int. J. Numer. Meth. Engng.} , vol.  88, 257–-278}

\hangindent=0.4cm\hangafter=1\noindent
{\color{black}{\bf Jackson,T.L.; Hooks, D.E.; Buckmaster, J.} (2011)
Modeling the microstructure of energetic materials with realistic constituent morphology.
{\it Propellants Explos. Pyrotech.}, vol. 36, pp. 252--258}

  \hangindent=0.4cm\hangafter=1\noindent
 {\bf  Khoroshun, L. P.} (1974): Prediction of thermoelastic properties of materials strengthened by
unidirectional discrete fibers. {\it Prikladnaya
Mekhanika}, vol. 10(12), pp. 23--30. (In  Russian. Engl. Transl. {\it Soviet Appl. Mech.}
 vol. 10, pp. 1288--1293)

\hangindent=0.4cm\hangafter=1\noindent
 {\bf Khoroshun, L.P.}  (1978): Random functions theory in problems on the macroscopic
characteristics of microinhomogeneous media. {\tenit Priklad Mekh}, vol. 14(2), pp.3--17
(In Russian. Engl Transl. {\tenit Soviet Appl Mech}, vol. 14, pp. 113--124)

\hangindent=0.4cm\hangafter=1\noindent
{\bf
Kreger, I.W. } (1972):  Rheology of monodisperse lattices. {\it  Adv Colloid and Interface
Sci},  vol.{ 3}, pp. 111--136

\hangindent=0.4cm\hangafter=1\noindent
{\bf
 Kr\"oner, E. } (1958):
 Berechnung der elastischen Konstanten des Vielkristalls
aus den Konstanstanten des Einkristalls.
{\it  Z. Physik.}  vol.{ 151}, pp. 504--518

\hangindent=0.4cm\hangafter=1\noindent
{\color{black} {\bf Kumar, N.C.; Matou\v{s}, K.; Geubelle, P.H.} (2008): Reconstruction of periodic unit cells of multimodal random particulate composites using genetic algorithms. {\it Computational Materials Science}, vol. 42, pp. 352--367}

\hangindent=0.4cm\hangafter=1\noindent{\bf
Kushch, V.I.; Sevostianov, I.; Mishnaevsky, L. } (2008):
Stress concentration and effective stiffness of aligned fiber reinforced composite with anisotropic constituents.
{\it  Int. J. Solids and Structures},  vol.{ 45}, pp. 5103--5117

\hangindent=0.4cm\hangafter=1\noindent{\bf
Lax, M. } (1952):  Multiple scattering of waves II. The effective fields dense systems.
{\it  Phys. Rev.},  vol.{ 85}, pp. 621--629

\hangindent=0.4cm\hangafter=1\noindent{\bf
Lee, C.H.; Gillman, A.S.; Matou\v{s}, K. } (2011):
Computing overall elastic constants of polydisperse
particulate composites from microtomographic data.
{\it  J. Mech. Phys. Solids},  vol.{ 59}, pp. 1838--1857

\hangindent=0.4cm\hangafter=1\noindent{\bf
Lee, H.; Brandyberry, M.; Tudor, A.; Matou\v{s}, K.}
 (2009): Three-dimensional reconstruction of statistically optimal unit cells of polydisperse particulate composites from microtomography. {\it Phys. Rev. E}. vol. 80, 061301 (12 pages)

\hangindent=0.4cm\hangafter=1\noindent{\bf
 Legrain,G.; Cartraud, P.;  Perreard, I.;  Mo\"es, N. } (2011):
An X-FEM and level set computational approach for image-based modelling: Application to homogenization
{\it  Int. J. Numerical Methods in Engineering}, vol. {
86}, pp. 915--934

\hangindent=0.4cm\hangafter=1\noindent{\bf
Lewandowski, J.J.; Liu, C.; Hunt Jr. W.H. } (1989):
Effects of matrix microstructure and particle distribution on fracture of an aluminum metal matrix composites.
{\it  Mater. Sci. Engineering},  vol.{ A 107}, pp. 241--255

\hangindent=0.4cm\hangafter=1\noindent{\bf
Liu, W.K.; Siad, L.; Tian, R.; Lee, S.;  Lee, D.,
Yin, X.; Chen, W.; Chan, S.; Olson, G.B.; Lindgen, L.-E.;
Horstemeyer, M.F.; Chang, Y.-S.; Choi, J.-B.; Kim, Y.J. } (2009):
Complexity science of multiscale materials via
stochastic computations. {\it  Int.  J. Numerical Methods in Engineering},  vol.{ 80}, pp. 932--978

\hangindent=0.4cm\hangafter=1\noindent{\color{black}{\bf
Liu, Y.; Greene, M.S.;  Chen, W.; Dikin, D.A.; Liu, W. K.} (2012):
Computational microstructure characterization and reconstruction for
stochastic multiscale material design. {\it Computer-Aided Design}. (In press).}

\hangindent=0.4cm\hangafter=1\noindent{\bf
Lusti, H.R.; Gusev, A.A.} (2004): Finite element predictions for the thermoelastic
properties of nanotube reinforced polymers. {\it Model. Simul. Mater. Sci. Eng.}, vol. 12,
pp. S107–-S119

\hangindent=0.4cm\hangafter=1\noindent{\bf
Maggi, F.; Bandera, A.; Galfetti, L.; De Luca, L.T.; Jackson, T. L. } (2010):
Efficient solid rocket propulsion for access to space.
{\it  Acta Astronautica},  vol.{ 66}, pp. 1563--1573

\hangindent=0.4cm\hangafter=1\noindent{\bf
Maggi, F.; Stafford, S.; Jackson, T.L.; Buckmaster, J. } (2008):  Nature of packs used in propellant modeling. {\it  Physical Review}  vol.{ E77}, 046107


\hangindent=0.4cm\hangafter=1\noindent{\bf
Massa, L.; Jackson, T.L.; Short, M. } (2003):
Numerical solution of three-dimensional
heterogeneous solid propellants
{\it  Combust. Theory Modeling},  vol.{ 7}, pp. 579--602

\hangindent=0.4cm\hangafter=1\noindent {\color{black}
{\bf  Massoni, J.; Saurel, R.; Baudin, G.; Demol, G.} (1999) A mechanistic model for shock initiation of solid explosives.
{\it Physics of Fluids}, vol. 11, pp. 710--736}

\hangindent=0.4cm\hangafter=1\noindent{\bf
Matou\v{s}, K.; Geubelle, P.H. } (2006):
Multiscale modelling of particle debonding in reinforced
elastomers subjected to finite deformations. {\it  Int.  J. Numerical Methods in Engineering},  vol.{ 65}, pp. 190--223

\hangindent=0.4cm\hangafter=1\noindent{\bf
Matou\v{s}, K.; Inglis H.M.; Gu X.; Rypl, D.; Jackson, T.L.; Geubelle, H.P. } (2007):
Multiscale modeling of solid propellants: From particle packing to failure.
{\it  Composites Science and Technology},  vol.{ 67}, pp. 1694--1708

\hangindent=0.4cm\hangafter=1\noindent{\color{black}{\bf
Matou\v{s}, K.; Lep\v{s}, M.; Zeman, J.;  \v{S}ejnoha M.} (2000): Applying genetic
algorithms to selected topics commonly encountered in engineering practice. {\it Computer
Methods in Applied Mechanics and Engineering}, vol. 190, pp. 1629--1650}

\hangindent=0.4cm\hangafter=1\noindent{\bf
Milton, G. W. } (2002): {\it  The Theory of Composites.} Appl. Comput. Math.;   vol.{ 6}, Cambridge University Press

\hangindent=0.4cm\hangafter=1\noindent{\bf
{Mori, T.}; {Tanaka, K.} } (1973):  Average stress in matrix and average elastic energy of materials with misfitting inclusions. {\it  Acta Metall}.  vol.{ 21}, pp. 571--574


\hangindent=0.4cm\hangafter=1\noindent{\bf
{Nemat-Nasser, S.}; {Hori, M.} } (1993):  {\it  Micromechanics: Overall Properties of Heterogeneous Materials}. Elsevier, North-Holland

\hangindent=0.4cm\hangafter=1\noindent{\bf
Patlazhan, S.A. } (1993):  Effective viscosity theory of a random concentrated suspension of polydisperse hard spheres. {\it  Physica}  vol.{ A198}, pp. 385--400

\hangindent=0.4cm\hangafter=1\noindent{\bf
Percus, J.K.; Yevick, G.J. } (1958):  Analysis of classical statistical mechanics by means
of collective coordinates. {\it  Phys. Rev.},  vol.{ 110}, pp. 1--13

\hangindent=0.4cm\hangafter=1\noindent
{\bf Ponte Casta\~neda, P.; Willis, J.R.} (1995): The effect of spatial distribution on the
effective behavior of composite materials and cracked media. {\it J. Mech. Phys.
Solids}, vol.  43, pp. 1919–-1951

\hangindent=0.4cm\hangafter=1\noindent{\color{black}{\bf
Povirk, G. L.} (1995): Incorporation of microstructural information into
models of two-phase materials, {\it Acta Metallurgica et Materialia}, vol. 43, pp. 3199--3206}

\hangindent=0.4cm\hangafter=1\noindent{\bf
Probstein, R.F.; Sengun, M.Z.; Tseng, T.-C. } (1994):  Bimodal model of concentrated suspension viscosity for
distributed particle sizes. {\it  J. Rheol.},  vol.{ 38}, pp. 811--829

\hangindent=0.4cm\hangafter=1\noindent{\bf
Ravichandran, G;  Liu, C.T} (1995):
Modeling constitutive behavior of particulate composites undergoing damage.
{\it Int. J. Solids and Structures}, vol. 32, pp. 979-–990

\hangindent=0.4cm\hangafter=1\noindent{\color{black}{\bf
Rayleigh, L.} (1892): On the influence of obstacles arranged in rectangular order upon the properties of a medium.
{\it Philosophical Magazine}, vol.  34, pp. 481--502}

\hangindent=0.4cm\hangafter=1\noindent{\bf
Segurado, J.; Llorca, J. } (2002):  A numerical approximation to the elastic properties of sphere-reinforced composites. {\it  J. Mech. Phys. Solids},  vol.{ 50}, pp. 2107--2121

\hangindent=0.4cm\hangafter=1\noindent{\bf
Shapiro, A.P.; Probstein, R.F. } (1992):  Random packings of spheres and fluidity limits of
monodisperse and bidisperse suspensions. {\it  Phys. Rev. Lett.},  vol.{ 68}, pp. 1422--1425

\hangindent=0.4cm\hangafter=1\noindent{\color{black}{\bf
Shermergor, T. D.} (1977): {\it The Elasticity Theory of Microheterogeneous Media.} Nauka, Moscow.
(in Russian)}

\hangindent=0.4cm\hangafter=1\noindent{\bf
Smith, J. } (1976):  The elastic constants of a particulate filled glassy polymer.
{\it  Polymer. Eng. Sci.},  vol.{ 16}, pp. 394--399

\hangindent=0.4cm\hangafter=1\noindent{\bf
Stickel, J.J.; Powell, R.L. } (2005):  Fluid mechanics and rheology of dense suspensions.
{\it  Annu. Rev. Fluid Mech.,}  vol.{ 37}, pp. 129--149

\hangindent=0.4cm\hangafter=1\noindent{\bf
Stroeven, M.;  Askes, H.; Sluys, L.J. } (2004):
Numerical determination of representative volumes for granular materials.
{\it  Comput. Methods Appl. Mech. Engrg.}  vol.{ 193}, pp. 3221--3238

\hangindent=0.4cm\hangafter=1\noindent{\bf
Sutton, G.P.; Biblarz, O.} (2003): {\it Rocket Propulsion Elements.}
John Wiley \& Sons, NY

\hangindent=0.4cm\hangafter=1\noindent{\bf
Sutton, M.A.; Orteu, J.J.; Schreier, H.W. } (2009):  {\it  Image Correlation for Shape, Motion and Deformation Measurements}.  Springer, NY

\hangindent=0.4cm\hangafter=1\noindent{\bf
Tan, H.; Huang, Y.; Liu, C.; Geubelle, P.H. } (2005):
The Mori--Tanaka method for composite materials with nonlinear interface
debounding. {\it  Int. J. Plasticity},  vol.{ 21}, pp. 1890--1918

\hangindent=0.4cm\hangafter=1\noindent{\bf
Tan, H.; Huang, Y.; Liu, C.; Inglis, H.M.; Ravichandran, G.; Geubelle, P.H.} (2007) The uniaxial tension of
particle-reinforced composite materials with nonlinear interface debonding. {\it Int. J. Solids Struct}. vol 44, pp. 1809--1822

 \hangindent=0.4cm\hangafter=1\noindent{\bf
Tawerghi, E.; Yi, Y.-B. } (2009):  A computational study on the effective properties
of heterogeneous random media containing particulate inclusions. {\it  J. Physics. D. Applied Physics},
 vol.{ 42}, pp. 175409

 \hangindent=0.4cm\hangafter=1\noindent{\bf
Torquato, S. } (2002):  {\it  Random Heterogeneous Materials:
Microstructure and Macroscopic Properties.}
Springer-Verlag

\hangindent=0.4cm\hangafter=1\noindent
{\bf Willis, J. R.}  (1977):
Bounds and self-consistent estimates for the overall properties
of anisotropic composites.
 {\it J. Mech. Phys. Solids},
 vol.  25,   pp. 185--203

\hangindent=0.4cm\hangafter=1\noindent{\bf
Willis, J. R. } (1978):  Variational principles and bounds for the overall properties
of composites. In: Provan JW (ed), {\it  Continuum Models of Disordered Systems}.
University of Waterloo Press, Waterloo, pp. 185---215

\hangindent=0.4cm\hangafter=1\noindent{\bf
Willis, J. R. } (1981):  Variational and related methods for the overall properties
of composites. {\it  Advances in Applied Mechanics},
 vol.{ 21}, pp. 1--78

\hangindent=0.4cm\hangafter=1\noindent{\bf
Wissler, M.; Lusti, H.R; Oberson, C.; Widmann-Schupak, A.H.; Zappini, G.; Gusev, A.A. } (2003):
Non-additive effects in the elastic behavior of dental composites.
{\it  Advanced Engineering Materials},  vol.{ 5}, pp. 113--116

\hangindent=0.4cm\hangafter=1\noindent{\bf
 Wriggers, P.; Moftah, S.O. } (2006):  Mesoscale models for concrete: Homogenisation and damage behavior.
{\it  Finite Elements in Analysis and Design},  vol.{ 42}, pp. 623--636

\hangindent=0.4cm\hangafter=1\noindent{\bf
Xu, F.; Aravas, N.; Sofronis, P. } (2008):
Constitutive modeling of solid propellant materials with
evolving microstructural damage.
{\it  J. Mechanics Physics Solids},  vol.{ 56}, pp. 2050--2073

\hangindent=0.4cm\hangafter=1\noindent
{\color{black}{\bf Yanovsky, V.E.; Zgaevskii, V.E.} (2004)
Mechanical properties of high elastic polymer matrix composites filled with rigid particles: Nanoscale consideration of the interfacial problem
{\it Composite Interfaces}, vol. 11, pp. 245-261}

\hangindent=0.4cm\hangafter=1\noindent{\bf
Zaman, A. A.; Moudgil, B. M. } (1998):
Rheology of bidisperse aqueous silica suspensions: A new scaling method for the bidisperse viscosity
{\it  J. Rheol.},  vol.{ 42}, pp. 21--39

\hangindent=0.4cm\hangafter=1\noindent
{\bf Zeman, J.; \v{S}ejnoha, M.} (2001): Numerical evaluation of {\color{black}effective} elastic properties of graphite fiber tow impregnated by polymer matrix. {\it J.
Mech. Physics of Solids}, vol. 49, pp. 69-–90

\hangindent=0.4cm\hangafter=1\noindent{\bf
Zhong, X.A.; Knaus, W.G. } (2000):  Effects of particle interaction and size variation
on damage evolution in filled elastomers. {\it  Mechanics Compos. Materials Structures},  vol.{ 7}, pp. 35--53

\hangindent=0.4cm\hangafter=1\noindent
{\color{black}{\bf Zgaevsky, V.E.} (1977):
Elastic and viscoelastic properties of polymers filled with solid particles.
{\it Int. J. Polym. Mater.},  vol. 6, pp. 109--124

}
\end{document}